# A peeling approach for integrated manufacturing of large mono-layer h-BN crystals

Ruizhi Wang[†], David G. Purdie[†,‡], Ye Fang[†], Fabien Massabuau[§], Philipp Braeuninger-Weimer[†], Oliver J. Burton[†], Raoul Blume[∥], Robert Schloegl[⊥], Antonio Lombardo[†,‡], Robert S. Weatherup[∇,o], Stephan Hofmann*,[†]

[†]Department of Engineering, University of Cambridge, 9 JJ Thomson Avenue, Cambridge CB3 0FA, United Kingdom

[‡]Cambridge Graphene Centre, University of Cambridge, 9 JJ Thomson Avenue, Cambridge CB3 0FA, United Kingdom

[§]Department of Materials Science and Metallurgy, University of Cambridge, 27 Charles Babbage Road, Cambridge CB3 0FA, United Kingdom

[∥]Helmholtz-Zentrum Berlin für Materialen und Energie, D-12489 Berlin, Germany

[⊥]Fritz Haber Institute, D-14195 Berlin-Dahlem, Germany

[∇]School of Chemistry, University of Manchester, Oxford Road, Manchester M13 9PL, UK

[o]University of Manchester Harwell Campus, Diamond Light Source, Didcot, Oxfordshire, OX11 0DE, UK

# Abstract

Hexagonal boron nitride (h-BN) is the only known material aside from graphite with a structure composed of simple, stable, non-corrugated atomically thin layers. While historically used as lubricant in powder form, h-BN layers have become particularly attractive as an ultimately thin insulator. Practically all emerging electronic and photonic device concepts rely on h-BN exfoliated from small bulk crystallites, which limits device dimensions and process scalability. Here, we address this integration challenge for mono-layer h-BN via a chemical vapour deposition process that enables crystal sizes exceeding 0.5 mm starting from commercial, reusable platinum foils, and in unison allows a delamination process for easy and clean layer transfer. We demonstrate sequential pick-up for the assembly of graphene/h-BN heterostructures with atomic layer precision, while minimizing interfacial contamination. Our process development builds on a systematic understanding of the underlying mechanisms. The approach can be readily combined with other layered materials and opens a scalable route to h-BN layer integration and reliable 2D material device layer stacks.

Scalable manufacture remains a central challenge in the application of two-dimensional layered materials (2DLMs). In recent years, major advances have been made regarding chemical vapour deposition (CVD) of 2DLMs such as graphene (Gr)[1,2] and hexagonal boron nitride (h-BN).[3,4] Many studies have revealed the details of the growth processes on select, generally catalytic, substrates.[5–7] The focus thereof has been on achieving ever-larger single-crystalline regions by lowering the nucleation density[8,9] and/or by merging aligned domains.[10,11] However, emerging applications require integration into device stacks. The most versatile route for this is 2DLM transfer from the CVD growth catalyst to the designated device, often including the vertical stacking of 2DLMs to form van der Waals heterostructures. A number of transfer methods have been proposed for CVD Gr and h-BN, including wet transfer,[12,13] dry transfer,[14–16] electrochemical delamination[17,18] and lift-off-transfer.[19] Progress in developing these methods has not kept pace with large-area 2DLM crystal growth and the introduction of contamination and damage remain major constraints, which is exceptionally severe in case of heterostructures that rely on atomically clean interfaces.[20,21]

Heterostructure devices fabricated entirely using exfoliated 2DLM now achieve carrier mobility values close to theoretically-predicted limits.[22–24] As exfoliation is an inherently non-scalable approach to fabrication, significant efforts have been made to replace each of the constituent layers using CVD 2DLMs in a step-wise manner.[25,26] The approach for high mobility Gr channels has been to use h-BN flakes exfoliated from bulk crystallites to peel off CVD Gr from the growth substrate, thus minimising transfer related defects and interface contamination. Using dry transfer, Cu catalysed CVD Gr has been shown to exhibit electron and hole mobilities well above 50,000 $cm^2$/Vs at room temperature[25] and ballistic transport lengths of 28 μm have been demonstrated at temperatures below 2 K, limited solely by device dimensions, i.e. the size of exfoliated h-BN flakes.[26] While highlighting the potential for applications particularly in optoelectronics and sensing, these examples still rely on mechanical exfoliation from bulk h-BN crystallites, giving limited, random flake size and varying thickness. Such dependence on flake exfoliation is a major bottleneck for further advances toward scalable device integration.

Here, we focus on addressing this challenge of reliably integrating CVD h-BN into 2DLM heterostructures for scalable manufacture. This demands new holistic process development, not only growth of large h-BN mono-layer crystals, but also their viable, clean transfer and device interfacing for which we take high-mobility Gr channels as application relevant model system to assess quality. We investigate h-BN growth and transfer in unison, in particular in the context of the choice of growth catalyst since both processes critically rely on h-BN interactions with the catalyst. Strong metal/h-BN interactions are preferable for applications where the h-BN remains on the metal, such as for magnetic tunnel junctions.[27,28] To enable effective transfer, however, only weak adhesion to the growth substrate is desired. Cu, which is widely employed as a catalyst for CVD Gr and h-BN,[7,12] exhibits a weak interaction with both materials.[29,30] The high vapour pressure of Cu at typical growth temperatures[31] leads to concerns not only about reactor contamination and a restricted high temperature parameter space, but also contamination of the 2DLM by trace Cu, which is a constraint for CMOS integration.[21] We therefore focus on Pt, which is also a weakly interacting catalyst[27] but with a lower vapour pressure and higher melting point.[32] h-BN interaction with Pt has been studied in detail for the Pt(111) surface.[33,34] Reported CVD h-BN domain sizes on Pt to date are typically only a few μm.[35,36] Much larger domain sizes have been reported for graphene CVD on Pt,[37] but in all cases 2DLM transfer relied on conventional electrochemical delamination.

Based on an understanding of the growth process, we present a CVD process to achieve large monolayer h-BN domains with lateral sizes exceeding 0.5 mm. Importantly we show that as-grown h-BN mono-layers can be easily and cleanly transferred using an entirely delamination-based approach, which also enables the reuse of the substrate. We demonstrate sequential pick-up to create h-BN films of controlled layer thicknesses and graphene/h-BN heterostructures, while minimizing interfacial contamination and showing high device performance. Our work opens a pathway to integrate CVD h-BN in high quality 2DLM heterostructures.

## Results

**Overview of growth process**

Fig. 1 gives an overview of our CVD process and how h-BN growth proceeds. In its basic form, our process, which we refer to as a sequential step growth (SSG), consists of two coupled borazine exposures at different pressures (Fig. 1a). In contrast, widely used single exposure at fixed temperature and pressure is referred to as "standard" growth (SG). We use borazine ($B_3H_6N_3$) as a combined boron (B) and nitrogen (N) precursor, which is isostructural to benzene and has a high vapour pressure of 340 mbar at room temperature.[38] In conjunction with our cold-wall CVD reactor system (base pressure $< 2 \times 10^{-6}$ mbar, further details in Experimental Methods), a well-controlled precursor atmosphere can be maintained, in contrast to the use of ammonia borane, which can exhibit a complex and evolving decomposition profile especially for hot-wall CVD reactor conditions.[39] Our main findings relate to fundamental interactions with the catalytic growth substrate and hence are transferable to other B and N precursors. In particular, we note that while borazine represents a precursor with pre-defined stoichiometry (B:N = 1), it dissociates on the Pt surface. This means that the interaction of the constituent elements with the catalyst will dictate the actual supply of the elements during CVD.[7] We use commercially available polycrystalline Pt foils (25 μm, 99.99%, Alfa Aesar) as growth substrates, which are remotely heated using an IR laser with a beam shaper that creates a homogeneous field with a top-hat profile. This minimises sample cross-contamination. The low thermal mass allows fast ramping to and from growth temperatures ($T_{gr}$) of up to 1300 °C. If not stated otherwise, all SEM images are taken immediately after growth with around 30 min or less atmospheric exposure for transfer. This is important as the secondary electron (SE) contrast for h-BN mono-layers can change on Pt, as we will discuss below. Further details are given in the Experimental Section.

The first exposure in SSG (Fig. 1a), at relatively high borazine pressure ($P_{sd}$), promotes recrystallization and grain growth of the Pt foil which is initially highly poly-crystalline (Fig. 1b). The h-BN nucleation density and homogeneity are then controlled by briefly removing the precursor during the homogenization stage ($t_{homo}$; see Fig. 1a), which leads to the dissolution and removal of excess h-BN. In the second exposure phase ($t_{exp}$) the nuclei are then laterally expanded into large mono-layer h-BN crystals via exposure at low borazine pressure ($P_{exp}$). Upon further exposure, these merge into a continuous h-BN mono-layer film. The h-BN mono-layers thus grown can then be delaminated directly from the Pt, which can be reused for further growth (see SI, Fig. S9).

**Investigation of growth mechanism using in-situ XPS**

It remains unclear from previous literature on Pt catalysed h-BN CVD whether growth occurs isothermally or via precipitation on cooling.[36] Hence, we make use of in-situ and ex-situ characterisation to establish a first order understanding of the underlying growth mechanisms as basis for further process development. In-situ X-ray photoelectron spectroscopy (XPS) provides surface-sensitive information on the growth mechanism of h-BN (see experimental section for details).[4,7] Fig. 2 shows in-situ XPS measurements of a Pt foil during a basic one-step borazine exposure (process diagram Fig. 2a). XPS measurements are taken throughout the CVD process, i.e. during heating to $T_{gr}$, subsequent borazine exposure at constant pressure $p_{sd}$ and cool-down. Fig. 2b shows the evolution of B1s and N1s core-level spectra taken during borazine exposure ($3\times10^{-4}$ mbar), at a $T_{gr}$ of 1100 °C (step II.1 – II.2). Both the B1s and N1s spectra show the emergence of a peak that grows in intensity with time, consistent with the isothermal growth of h-BN on the Pt surface. XPS spectra measured toward the end of borazine exposure show a main B1s peak centred at a binding energy of ~191.5 eV and N1s peak centred at ~399.0 eV (Fig. 2c). We observe the π→π* plasmon shake up satellite at ~200 eV corresponding to $sp^2$ bonded h-BN, which can be more clearly discerned when multiple spectra are summed to improve the signal to noise ratio (see SI, Fig. S2). Analysis of the relative peak intensities also confirms that the B:N ratio is ~1. Based on ex-situ measurements using transmission electron microscopy (TEM) and Raman spectroscopy (see below and SI), we exclude significant formation of cubic BN and multilayer h-BN. Hence we assign the given XPS signatures to mono-layer h-BN, consistent with previous literature.[7,30,40]

After the first exposure step in SSG and when removing the precursor (temperature kept at $T_{gr}$ = 1100 °C), the B1s and N1s peaks disappear within less than 2 minutes (Fig. 2c). The B1s and N1s peaks do not reappear upon cooling at these conditions. There are multiple possible explanations for the disappearance of the Pt supported h-BN mono-layer. One process that has been reported in the context of removing Gr grown on Pt is etching with $H_2$ at 1060 °C.[41] Other studies have investigated the stability of h-BN on $SiO_2$, a non-catalytic substrate, in an oxygen containing atmosphere and found the onset of degradation at 850 °C[42] and complete removal at 1000 °C.[43] The stability of h-BN may be further reduced when on a catalyst such as the growth substrate, as in the case of h-BN on Cu where depending on the state of oxygen intercalation, the h-BN can dissociate completely at 700 °C.[7] Given the absence of any gas after precursor removal and a base pressure of below 1 x 10$^{-8}$ mbar, we exclude the possibility of significant etching for our experiments. Instead, we suggest that at the given growth temperatures h-BN is not stable on the growth catalyst once the precursor has been removed, and either desorbs, is dissolved into the bulk or both. The bulk solubility of B in Pt is reported to be ~2.5 at% at $T_{gr}$ = 1100 °C.[44] We could not find any relevant ternary phase diagrams in literature, or reliable data on N solubility in Pt, but it has previously been assumed that the N solubility in bulk Pt is negligible.[35] Simple thermodynamic bulk solubility considerations are typically inadequate for describing growth of nanomaterials where kinetic processes and local supersaturations are often dominant.[45] Furthermore, under certain conditions (see SI, Fig. S4) we have observed precipitation effects indicating not only boron (see also Fig. 3) but also nitrogen solubility, consistent with the h-BN layers dissolving into the catalyst bulk (Fig. 2c).

As well as revealing the presence and evolution of h-BN on the Pt surface, we are also able to monitor the chemical state of the Pt catalyst with in situ XPS. Fig. 2d presents the Pt 4f core level spectra for bare (prior to borazine exposure) and h-BN covered Pt surfaces at two different X-ray excitation energies taken at $T_{gr}$ = 950 °C. Photoelectrons collected with higher X-ray energy have longer inelastic mean free paths ($\lambda_{IMFP}$ ≈ 4.5 Å at 225 eV, $\lambda_{IMFP}$ ≈ 7.6 Å at 525eV),[46] and thus higher information depth, i.e. the data represents depth resolved information about the catalyst surface composition. We observed no changes in the Pt4f spectrum compared to the clean surface indicating the absence of any significant B and N phases near the surface.

The results for CVD of h-BN on Pt are consistent with a first-order kinetic growth model previously introduced for other transition metal catalysts, which takes into account precursor and elemental flux balances.[7,45] The key findings regarding the optimized CVD process in Fig.1 are that h-BN grows isothermally on Pt and no further growth occurs upon cooling due to precipitation. This is in contrast h-BN growth on Fe, where the contribution to growth due to precipitation is significant and thus control thereof is crucial for growth of large grain size h-BN.[4] At the chosen CVD conditions, h-BN growth appears to be driven by a local B and N supersaturation at the surface. At high temperatures ($T_{gr}$ = 1100 °C, Fig. 2c) the Pt supported h-BN mono-layer film dissociates and at least partly dissolves into the Pt bulk once the borazine flux is removed. Within the parameter space covered by our in- and ex-situ experiments (max. borazine pressure of $10^{-2}$ mbar; max. temperature of 1400°C; max. exposure time of 30 min) we only observe monolayer h-BN on Pt. This is independent of the cooling rate which we have varied from 10 °C/min to immediate quenching (temperature drop of >500°C in a period of ~5s). This is in contrast to h-BN growth on Fe, where multi-layer h-BN formation is observed.[4]

**Details of growth process**

Fig. 3 shows the effect of annealing of the Pt foil and how the choice of CVD atmosphere prior to growth impacts its texture and crystallinity. h-BN growth is linked to the nature of the growth substrate, analogous to graphene CVD. Different catalyst foil facets show different catalytic growth activities and lead to varying 2DLM nucleation density, domain alignment and domain shape evolution.[47] Hence optimising the crystallinity of the catalyst foil is an important part of the CVD process. A first indication of the crystallinity can be obtained through SEM. Crystal domains with different orientations will differ in brightness due to channelling contrast. The lack thereof gives a strong indication of a single Pt crystal facet. We carried out a systematic set of experiments with commercial Pt foils that were annealed in vacuum for 15 min, similar to the general SSG process outlined in Fig. 1, and then subjected to an additional 2 min of annealing under different gas environments (Fig. 3a). For annealing in vacuum (at base pressure < 2 x $10^{-6}$ mbar), $H_2$ ($10^{-3}$ mbar) or $NH_3$ ($10^{-3}$ mbar), SEM shows a binary distribution of grain sizes. The majority of the Pt grains have a lateral size around 50 μm – 100 μm (Fig. 3b, Loc. 1). A few selected Pt grains grow significantly larger, reaching millimetre-scale (Fig. 3b, Loc. 2). In contrast, when annealed in borazine for 2 min, at otherwise identical conditions, the Pt surface shows uniform SEM contrast across a few mm$^2$ of sample area, indicative of a dominant single in-plane orientation (Fig. 3b). No evidence of h-BN growth is observed at this stage. X-ray diffraction measurements (XRD) were taken to investigate the Pt foil texture in more detail after borazine annealing. The 2θ-ω scan presented in Fig. 3c shows a dominant Pt (1 1 1) orientation following annealing. This is to be expected given that (1 1 1) is the lowest energy interface for FCC crystals, such as Pt. The texture map in Fig. 3d of the Pt (1 1 1) reflection (at 2θ = 39.73º) shows one pole in the symmetric position (χ ~ 0º) and 3 poles at χ ~ 70° and spaced by ϕ = 120° from each other. This is evidence that the vast majority of the Pt grains have the same orientation, i.e. the grains are not rotated relative to each other. The foil is not completely single-crystalline as highlighted by the additional weaker poles, for example the one at χ ~ 70º and ϕ = 90°, which indicate that there are some grains, albeit a minority (potentially in the bulk of the foil), that are rotated with respect to the dominant orientation. We note that the Pt surface roughness is not significantly altered by this crystallisation, with an average roughness measured by atomic force microscopy (AFM) of $R_a$ = 5.3 nm for the poly-crystalline and $R_a$ = 4.7 nm for the crystallized foil (see SI, Fig. S11).

There are a number of reports on thermal recrystallization of Cu foils for improved graphene CVD, using extended thermal pre-treatment[48,49] and thermal gradients.[11] We also explore thermal gradients (see SI, Fig. S1), and emphasize that longer annealing times lead to improved foil crystallisation. Here, however, we focus on the accelerated crystallisation observed for borazine exposure which has benefits as part of an integrated process. Upon thermal annealing, recrystallization is at first driven by a reduction of dislocation energy, followed by grain growth driven by the minimisation of the overall energy associated with grain-boundaries (GB). The latter involves normal grain growth where all grains grow uniformly, as well as abnormal grain growth (also known as secondary crystallization) where one type of grain will grow significantly faster than the others driven by the difference in surface energies between different grain orientations.[50–54] Our data shows that such abnormal growth in Pt is significantly enhanced by the presence of borazine. For Pt thin films it has been frequently observed that the presence of oxygen is detrimental to the formation of grains with (1 1 1) surface orientation.[51,55] In this context B, which can readily adsorb in Pt grain boundaries,[56] is frequently used as deoxidizer,[57] and the addition of B to Pt has been seen to cause GB unpinning.[54] We therefore attribute the accelerated abnormal Pt grain growth to GB unpinning via removal of pre-existing solutes. Another potential mechanism could be B causing solute drag by decorating and pinning selective GBs, leaving only high mobility GBs able to grow.

Having established the rationale for the catalyst foil crystallisation step, we now consider each subsequent step in the SSG process (Fig. 1a), i.e. seeding, homogenization and domain expansion. We carried out a systematic set of SG experiments and focus on the role of key parameters, particularly exposure times ($t_{sd}$, $t_{homo}$, $t_{exp}$), exposure pressures ($P_{sd}$, $P_{exp}$) and temperatures ($T_{gr}$), for each individual step of the integrated SSG process. In Fig. 4, a series of SEM images show the outcome of SG growth experiments at precursor pressures ($P_{sd}=1\times10^{-5}$ mbar) corresponding to those used in the first SSG step (Fig. 1a). The topmost images in Fig. 4b and 4c represent the outcome of seeding for $t_{sd}$ = 3 min and $t_{sd}$ = 5 min prior to homogenization (i.e. $t_{homo}$ = 0 min). A more detailed series is given in Fig. S5 in the SI. For $t_{sd}$ = 3 min a few h-BN domains of very similar sizes have nucleated. At $t_{sd}$ = 5 min these h-BN domains have grown in size, but secondary h-BN nucleation has occurred as evidenced by the many additional smaller domains. We note that the onset of secondary nucleation varies and starts occurring from $t_{sd}$ = 3min onwards. This highlights a key challenge for controlling the microstructure of the resulting h-BN. A high precursor exposure pressure is desirable to minimise incubation time and achieve rapid nucleation and growth (compare to Fig. S6 in SI for SG at lower $P_{sd}$), however the accompanying secondary nucleation is detrimental to achieving large domain sizes. We thus introduce a homogenization stage in SSG, and figs. 4b and 4c show the effects of increasing $t_{homo}$. After $t_{homo}$ = 5 min all smaller h-BN secondary nuclei have disappeared (Fig. 4c) and for $t_{homo}$ = 10 min also the larger h-BN domains are removed or significantly reduced in size. Our XPS data (Fig. 2) showed that existing h-BN nuclei are unstable and start to dissociate when borazine is removed. Given a constant rate of dissociation, it is expected that smaller nuclei will disappear first, consistent with our data. This allows us to control the h-BN nucleation density, which justifies our choice $t_{homo}$ = 5 min to achieve optimal growth results (Fig. 1a). It should be noted that the overall parameters are highly interdependent. Thus the suggested time of $t_{homo}$ = 5 min is optimized for the given set of parameters of temperature, precursor pressure, seeding time and catalyst dimensions. Following homogenization (see Fig. 1a, IV-VII) we then implement a domain growth stage at lower borazine pressure of $2.5 \times 10^{-6}$ mbar. Through this separation of nucleation and domain growth, we can effectively supress secondary nucleation and thus achieve extremely large h-BN crystal sizes.

**Characterization of h-BN crystal alignment**

A combination of characterization techniques is used to assess the quality of the h-BN films. Fig. 5a shows a representative bright field (BF) transmission electron microscopy (TEM) image. The h-BN film is only indirectly visible by the presence of a suspended particle (see dotted circle, Fig. 5a), which highlights that the h-BN is uniform and has little contrast and/or features. The selected area electron diffraction pattern in dark field (DF) TEM (Fig. 5b) shows sets of hexagonal diffraction patterns consistent with single crystal mono-layer h-BN. To determine the crystal orientation over a reasonably large area, DF-TEM diffraction patterns from various points across the h-BN film were recorded and analysed. We define α as the angle between the vertical and the closest first-order diffraction spot in a clockwise direction (Fig. 5b). The resulting distribution of α is summarised in Figs. 5c and d. For most recorded diffraction patterns α lies within a margin of ±2.5° of the median ($α_{med}$). The peaks at -30° and 30° correspond to identical orientations due to the hexagonal lattice symmetry. This distribution of orientations is in good agreement with previous studies,[58] and highlights that the h-BN is highly crystalline. Over the mapped area (~2 × 2 mm) the TEM signature is as expected for a single crystal, although we cannot rule out small rotations below the limit of resolution or defects induced by imperfect merging of domains. Figs. 5e, f show the edge of the film, with only one fringe visible, consistent with mono-layer h-BN. The region marked with the white arrow in Fig. 5f corresponds to a fold. The contrast between the h-BN mono-layer edge and folded edge can be clearly seen. We further assessed the layer number and homogeneity of as-grown h-BN through SEM and AFM. In all SEM images, where h-BN was immediately imaged, the h-BN domains have homogenous contrast, which indicates a constant number of layers across the whole observed region. AFM measurements show a ~0.4 nm step height for h-BN after transfer onto $SiO_2$ (see SI, Fig. S10), consistent with mono-layer h-BN thickness.

**Peeling transfer**

The SEM images of h-BN flakes shown so far (Figs. 1 and 4), were taken immediately after growth (reactor to SEM transfer time of around 30 min or less). Fig. 6a shows a SEM image recorded 5 hours after removing the sample from the growth chamber, during which time it was stored in ambient environment. Unlike the SEM immediately post-growth of an h-BN domain of similar size and grown under identical conditions (see Fig. 1b V), where the h-BN appears uniformly darker relative to the Pt(111) substrate, in Fig. 6a the contrast of the h-BN domain is not uniform. The observed lighter edge region (marked "Decoupled" in Fig. 6a) and contrast change is triggered by the intercalation of oxygen species. This has been previously observed in SEM and for instance LEEM for various systems of 2DLM on weakly interacting substrates.[27,59,60] The change takes place at room temperature within the time scale of hours (see SI, Fig S12), similar to G/Pt[27] and G/Cu.[61] The fact that intercalation occurs uniformly from the edges in Figs. 6a and S12 is a further indication of the quality of the h-BN, as the presence of large defects (such as grain boundaries or pinholes) within a h-BN domain would reveal itself by the onset of local intercalation. In fact, even for domains that have partially merged, we observe intercalation to only proceed from the edges and not from where any potential GB would be located (see SI, Fig. S13).

The observed intercalation at the interface between h-BN and Pt is an indication of their weak interaction, and a key motivation of our work. Here, we introduce a dry transfer approach to transfer h-BN grown on Pt, as schematically shown in Fig. 6b. A polyvinyl acetate (PVA) stamp is applied to the as-grown h-BN film through drop-casting. The PVA/h-BN stack is then delaminated mechanically (Fig. 6b I). For monolayer h-BN transfer this stack is then stamped down onto the target substrate (Fig. 6b IV) and the stamp is removed by dissolution in water. In line with previous experiments on delamination of Gr from Cu,[16,25] we observe an improvement in the ease of transfer by leaving the sample in an ambient environment for an extended period (typically >24h). We relate this effect to the decoupling of the h-BN layer, consistent with the time dependent change in SE contrast of h-BN on Pt (Fig. 6a; Fig S12 & S13 in SI).

The given method of dry transfer is not limited to monolayer transfer, but can also be used for the assembly of 2DLM stacks. After picking up the first h-BN layer, the same PVA/h-BN stack can be used for repeated exfoliation of further 2DLM layers. This is performed by simply stamping the stack onto another as-grown 2DLM on its growth substrate (Fig. 6b II) followed by mechanical delamination (Fig. 6b III), which results in the pick-up of an additional 2DLM layer. Such an approach seeks to keep interfacial contamination to a minimum as the second 2DLM layer is only ever in contact with the growth catalyst and the first 2DLM layer. Through sequential pickup in this way, it is possible to assemble stacks of 2DLMs. This approach is not limited to a specific 2DLMs. However, it requires the adhesion between stamp and 2DML to be higher than between 2DLM and substrate, highlighting the need for weakly interacting substrates. We use the sequential pick up of 2DLM to fabricate a variety of structures, with Fig. 6c (left) showing an optical image of multilayer h-BN on $SiO_2$ obtained by four sequential peelings of CVD mono-layer h-BN. A heterostructure of CVD h-BN and CVD Gr grown on Cu (Gr between h-BN and $SiO_2$) was also fabricated and an optical image of the result is shown in Fig. 6c (right). Since the transfer does not involve any modification of the growth catalyst, the Pt foil can be reused for additional growth cycles. In fact, Pt substrates were used for multiple growth runs (see SI, Fig. S9).

In order to assess the quality of the stacks and to confirm their structure, we employ Raman spectroscopy. The Raman spectrum of h-BN is characterised by the $E_{2g}$ peak at ~1370 cm$^{-1}$, which due to its non-resonant nature is very weak.[62] However, its peak position and relative intensity can offer an insight into the thickness of the h-BN.[62–64] Fig. 6d shows such Raman characterization of multilayer h-BN obtained through repeated exfoliation. The Raman spectra given in Fig. 6d (left) show an increase in $E_{2g}$ peak area, which is proportional to the number of sequential transfers. This demonstrates that using our transfer method we are able to pick-up an additional layer during each cycle. Furthermore, it should be noted that the full-width at half maximum (FWHM) of the $E_{2g}$ peak of monolayer h-BN is ~13cm$^{-1}$, which is comparable to exfoliated h-BN.[62] Since the FWHM is often cited as an indicator of material quality, it highlights the high quality of our CVD h-BN.[64] Fig. 6d (right) reflects the results of Raman mapping across 100 μm x 100 μm, whereby the quantitative analysis is based on fitting Lorentzian curves to the $E_{2g}$ peak. The linear relationship between peak area and layer number is maintained across the mapped region, indicating the quality of the transfer method. Furthermore, in agreement with previous studies,[63] we observe a slight red-shift in the $E_{2g}$ peak position with increasing layer number from 1369.7cm$^{-1}$ for monolayer h-BN to 1368.7cm$^{-1}$ for four-layer h-BN, which indicates a clean interface as this shift relates to the interaction between layers.[62,63] The Raman characterization of h-BN/Gr heterostructures is more challenging, due to the large difference in signal intensity of Gr and h-BN. The Raman spectrum of an all CVD heterostructure is shown in Fig. 6e. Due to the proximity of peaks, minor imperfections in the Gr will result in the Gr D-peak (1350cm$^{-1}$)[65] overlapping with and potentially swamping the h-BN peak (1370cm$^{-1}$). In Fig. 6e the Gr D-peak/G-peak ratio is less than 0.025, still the h-BN peak is only just visible in the magnification of the plot (inset), demonstrating that Gr can be directly delaminated from the growth catalyst using h-BN.

**Integration of CVD h-BN in Gr/h-BN heterostructures**

The fabrication of high quality Gr/h-BN heterostructures has been a significant challenge even when relying entirely on exfoliation.[23,66] Thus, to demonstrate the feasibility of our approach and to narrow-down the parameter space for device processing, we focus here on just the insertion of a CVD h-BN layer via a model structure consisting of monolayer CVD-h-BN as a capping layer on monolayer exfoliated (exf) Gr. Fig. 7a shows an optical image of the assembled stack. Using a PVA/CVD h-BN stack, which we obtain by delaminating an as grown h-BN layer from Pt, we pick up exf-Gr from a $SiO_2$/Si wafer. The h-BN is not observed optically as it uniformly covers the sample. It becomes apparent in Fig. 7a II and III, which presents the result of a peak force (PF-) AFM measurement of the transition region from CVD h-BN/exf-Gr to CVD h-BN only. The measured step height for the exf-Gr layer is ~0.4 nm and the surface is atomically smooth, which are indications of an interface without significant amounts of trapped residues. Fig. 7b shows the result of the Raman analysis. The Gr G peak and 2D peak position of each spectrum is plotted with the strain and doping axis for reference.[67] The colour of each point indicates the FWHM of the Gr 2D peak. The median of the Gr G peak and 2D peak positions are 1584.7 $cm^{-1}$ and 2684.6 $cm^{-1}$ respectively, compared to 1581.6 $cm^{-1}$ and 2676.9 $cm^{-1}$ for suspended exf-Gr.[67] The median FWHM of the Gr 2D peak is 26.2 $cm^{-1}$, which is similar to previous studies, where exf-h-BN/exf-Gr on $SiO_2$ has a 2D peak FWHM of about 25 $cm^{-1}$.[68] The measurement points are aligned along the strain axis, indicating that the sample is undoped, but is affected by strain.

The sheet conductivity of a representative CVD h-BN/exf-Gr field effect transistor (FET) device (where the Si substrate acts a back gate) is shown in Fig. 7c. The position of the charge neutrality point (CNP), at -0.2V, confirms that the graphene is highly intrinsic. The devices are shaped as Hall bars, allowing independent measurement of gate-dependent conductivity and charge carrier concentration (by applying an out of plane magnetic field). The Hall mobility ($\mu_H$) is then extracted assuming a Drude model for conductivity, leading to a peak $\mu_H$ = 7200 $cm^2V^{-1}s^{-1}$ at room temperature. The charge carrier density without the application of a back-gate voltage is $n_0 = 4.8 \times 10^{10}$ cm$^{-2}$, confirming the low doping level indicated by the Raman analysis. We note that such low doping is achieved without the need for any high temperature annealing to remove residuals post-transfer. We repeated the analysis for three independent devices and find the values reproducible. The performance of the given devices underscores the cleanliness of the graphene/h-BN interface and is, to the best of our knowledge, the highest for heterostructures using CVD h-BN. Further improvements in mobility should be achievable by introducing an additional h-BN layer underneath the Gr to screen the roughness and charged impurities of the SiO$_2$.[66] The given stack serves as a first demonstrator for our proposed approach of growing h-BN via CVD and integrating it into heterostructure devices with the goal to replace exfoliated h-BN.

## Conclusions

Improving current 2DLM integration strategies requires careful consideration of the growth catalyst. Here, we demonstrate that Pt is a suitable weakly interacting catalyst that enables not only h-BN growth with mono-layer control and crystal sizes in excess of 0.5 mm, but also direct transfer by delaminating as-grown layers to achieve clean processing. We establish that h-BN grows isothermally on Pt within the given CVD parameter space, driven by a local B and N supersaturation at the surface, and that when the precursor flux is removed the h-BN layer starts to dissolve. We find that initial B dissolution and decoration of Pt GBs leads to significantly accelerated Pt crystallisation. These insights allowed us to devise an integrated CVD growth process, that we referred to as SSG, which based on two coupled borazine exposures not only transforms poly-crystalline Pt foils into a dominant (111) orientation but enables independent control of h-BN nucleation and domain expansion, i.e. the CVD of highly crystalline mono-layer h-BN. We have demonstrated a delamination transfer method that makes use of the weak Pt/h-BN interaction, to directly delaminate the as-grown films from the catalyst. We thereby preserve the catalyst for regrowth and can achieve a clean transfer process. This approach allows the precise control of the thickness of the h-BN layer, something that has not been achieved for exfoliated h-BN, but is critical to many applications. Scalable manufacture, especially of clean 2DLM heterostructures, has been a main bottleneck for the whole field. While further process optimization is still required, our study opens a viable pathway towards achieving this elusive goal.

# Experimental Methods

**h-BN Growth**

All samples are grown in a custom-built Laser CVD reactor. An 808 nm continuous wave (CW) IR Laser with a maximum power of 60W is used for heating. It is positioned outside of the CVD chamber. The light is coupled into the chamber through a laser window. In order to achieve a homogeneous beam profile, a beam shaper is used that creates a top-hat (instead of Gaussian) profile at the focal point, which has a size of 5 mm x 5 mm. Due to very localized heating, only the sample itself is significantly heated, thus the system under discussion falls in the category of cold-wall reactors. We measure the temperature using an IR pyrometer with a wavelength of 1.6 μm, a focal spot diameter of 3 mm and the emissivity is set at 0.25.[69] The estimated uncertainty is ±50 °C.

If not specified otherwise, the Pt foil (25 μm, 99.99%, Alfa Aesar) is mounted on a tantalum (Ta) foil (25 μm, 99.9%, Goodfellow), which acts as the susceptor, enabling homogenous heating across a larger area than the size of the beam at the focal point. The Ta susceptor is clamped using sapphire in a custom-made mounting stage, and the Pt foil does not come into contact with anything but the susceptor, thus avoiding contamination and thermal dissipation. After loading the sample into the chamber, it is pumped down to a base pressure of less than $2 \times 10^{-6}$ mbar before starting growth. The pressure is measured using a full range pressure gauge consisting of Pirani and cold cathode gauges for different pressure regimes. All gases except borazine are injected into the chamber using mass flow controllers. The flow of borazine (>97%, Fluorochem) is controlled using a manual leak valve.

**Transfer**

**Peeling transfer** As the carrier layer for exfoliating the h-BN post-growth, we use a solution of 5g PVA (Mw 9000-10000, 80% hydrolized, Sigma Aldrich) and 1g glycerol as a softener (>99%, Sigmal Aldrich) in 100ml of de-ionized (DI) water. In the first step, we drop-cast the solution on the sample and dry at 80ºC for 20 min. Then, the stamp/h-BN film is peeled off using tweezers, stamped onto the target substrate at 125ºC, and annealed for 5 min. In the case of multilayer h-BN or h-BN/graphene heterostructures, the target substrate is another CVD grown 2DLM sample and the peeling process is repeated. In the final step, the stamp with the 2DLM layer is put down onto the substrate of choice and the carrier layer is dissolved in DI water at 50ºC for at least 3h.

**Bubbling transfer** The samples are spin-coated with Poly(methyl methacrylate) (PMMA, 495k) at 3000rpm for 40s then baked at 180 ºC for 1.5 min During the actual transfer process, NaOH solution (1M, a.q.) is used as electrolyte, and the sample (h-BN coated with PMMA) is used as the negative electrode and a Pt wire as the positive electrode. The typical settings are ~4.5V and ~0.3A. PMMA supported h-BN fully delaminates from Pt within 5 minutes. The sample is then rinsed in DI water three times, with each rinse lasting ~45 minutes, and then transferred onto the desired substrate.

**Characterization**

Raman measurements were performed with a Renishaw inVia confocal Raman Microscope. 514 nm of 532 nm lasers were used depending on equipment availability. Spectra were taken with a 50x objective lens. A step size of 2 μm was used for all maps. The SEM images were taken in the FEI Magellan SEM using an acceleration voltage of 1 kV. The TEM images (FEI

Osiris TEM) were taken with an acceleration voltage of 40kV. The knock-on damage of h-BN is minimized under such a low acceleration voltage. The samples are transferred using bubbling transfer in this case, as direct peeling requires PVA dissolution in DI water to release the h-BN film and the water's high surface tension is likely to damage the suspended h-BN. For the bubbling transfer only low surface tension solvents are used to release the h-BN.

In-situ XPS measurements were performed at the BESSY II synchrotron at the ISISS end station of the FHI-MPG. The setup consists of a reaction cell (base pressure $\approx 10^{-8}$ mbar) attached to an analyzer with a differentially pumped electrostatic lens system (Phoibos 150 NAP, SPECS GmbH).[70] XP core level spectra were collected in normal emission geometry using a x-ray beam with a spot size of ~80 μm × 150 μm. All spectra are background-corrected (linear) and their binding energies are referenced to the contemporaneously measured Fermi edges. The temperature is measured using a dual-wavelength pyrometer.

AFM was measured in peak force tapping mode using a Bruker Dimension Icon AFM. In this mode, the feedback loop keeps the peak force of tip-sample interaction constant.

XRD was carried out on a Philips X'pert MRD diffractometer with a Cu $K_{\alpha 1}$ X-ray source ($\lambda$ = 1.5405974 Å) and a 4-bounce Ge(220) asymmetric monochromator. The spot size is 5 mm x 15 mm.

**Heterostructure assembly, Device Fabrication & Measurement**

Continuous monolayer CVD h-BN on Pt was picked up/delaminated from Pt using the peeling procedure described above. The sample is then used to pick up Gr, which was prepared in advance by mechanical exfoliation from a bulk crystal onto a $SiO_2$/Si wafer. For heterostructure assembly, the PVA/h-BN layer is pressed onto the wafer with Gr at 30 °C and then peeled off.

This stack is then stamped onto another wafer at a temperature of 130 ºC. The sample is heated at 130 ºC for 5 min, before dissolving the PVA film in DI water.

In order to probe the electronic transport properties of the heterostructure we fabricate four-terminal transport geometries following an established method.[22,24] We begin by deposition of an aluminium (Al) etch mask fabricated by electron-beam (e-beam) lithography, followed by thermal evaporation of 30 nm of Al, and lift-off. Following this, the exposed regions of Gr/h-BN are etched with a reactive ion etcher (RIE) using $CF_4$ gas under a forward RF power of 20W. The Al mask is then removed by wet etching. Finally, metal contact leads are deposited by patterning with e-beam lithography followed by DC sputtering of 5/70 nm of Cr/Cu and lift-off.

Four-terminal transport measurements are performed in a Lakeshore Cryogenic probe station at a pressure of $\sim 4 \times 10^{-8}$ Torr and a temperature T=290K. The resistance is measured using a dual lock-in amplifier set-up at a frequency of ~13Hz and bias current ~100nA. The gate voltage is swept using an SMU.

# References

1. Ten years in two dimensions. *Ten years two Dimens. Nat. Nanotechnol. [Special Issue]* **9,** (2014).

2. Hofmann, S., Braeuninger-Weimer, P. & Weatherup, R. S. CVD-enabled graphene manufacture and technology. *J. Phys. Chem. Lett.* **6,** 2714–2721 (2015).

3. Pakdel, A., Bando, Y. & Golberg, D. Nano boron nitride flatland. *Chem. Soc. Rev.* **43,** 934–959 (2014).

4. Caneva, S. *et al.* Controlling catalyst bulk reservoir effects for monolayer hexagonal boron nitride CVD. *Nano Lett.* **16,** 1250–1261 (2016).

5. Loginova, E., Bartelt, N. C., Feibelman, P. J. & McCarty, K. F. Factors influencing graphene growth on metal surfaces. *New J. Phys.* **11,** 63046 (2009).

6. Bartelt, N. C. & McCarty, K. F. Graphene growth on metal surfaces. *MRS Bull.* **37,** 1158–1165 (2012).

7. Kidambi, P. R. *et al.* In situ observations during chemical vapor deposition of hexagonal boron nitride on polycrystalline copper. *Chem. Mater.* **26,** 6380–6392 (2014).

8. Wu, T. *et al.* Fast growth of inch-sized single-crystalline graphene from a controlled single nucleus on Cu-Ni alloys. *Nat. Mater.* **15,** 43–47 (2016).

9. Braeuninger-Weimer, P., Brennan, B., Pollard, A. J. & Hofmann, S. Understanding and controlling Cu-catalyzed graphene nucleation: the role of impurities, roughness, and oxygen scavenging. *Chem. Mater.* **28,** 8905–8915 (2016).

10. Lee, J.-H. *et al.* Wafer-scale growth of single-crystal monolayer graphene on reusable

hydrogen-terminated germanium. *Science (80-. ).* **344,** 286–289 (2014).

11. Xu, X. *et al.* Ultrafast epitaxial growth of metre-sized single-crystal graphene on industrial Cu foil. *Sci. Bull.* **62,** 1074–1080 (2017).

12. Li, X. S. *et al.* Large-area synthesis of high-quality and uniform graphene films on copper foils. *Science (80-. ).* **324,** 1312–1314 (2009).

13. Kim, K. S. *et al.* Large-scale pattern growth of graphene films for stretchable transparent electrodes. *Nature* **457,** 706–710 (2009).

14. Yoon, T. *et al.* Direct measurement of adhesion energy of monolayer graphene as-grown on copper and its application to renewable transfer process. *Nano Lett.* **12,** 1448–1452 (2012).

15. Bae, S.-H. *et al.* Unveiling the carrier transport mechanism in epitaxial graphene for forming wafer-scale, single-domain graphene. *Proc. Natl. Acad. Sci.* **114,** 4082–4086 (2017).

16. Whelan, P. R. *et al.* Raman spectral indicators of catalyst decoupling for transfer of CVD grown 2D materials. *Carbon N. Y.* **117,** 75–81 (2017).

17. Wang, Y. *et al.* Electrochemical delamination of CVD-grown graphene film: toward the recyclable use of copper catalyst. *ACS Nano* **5,** 9927–9933 (2011).

18. Liu, L. *et al.* A mechanism for highly efficient electrochemical bubbling delamination of CVD-grown graphene from metal substrates. *Adv. Mater. Interfaces* **3,** 150049 (2015).

19. Wang, R. *et al.* Catalyst interface engineering for improved 2D film lift-off and transfer. *ACS Appl. Mater. Interfaces* (2016). doi:10.1021/acsami.6b11685

20. Pirkle, A. *et al.* The effect of chemical residues on the physical and electrical properties of chemical vapor deposited graphene transferred to $SiO_2$. *Appl. Phys. Lett.* **99,** 122108 (2011).

21. Lupina, G. *et al.* Residual metallic contamination of transferred chemical vapor deposited graphene. *ACS Nano* **9,** 4776–4785 (2015).

22. Wang, L. *et al.* One-dimensional electrical contact to a two-dimensional material. *Science (80-. ).* **342,** 614–617 (2013).

23. Pizzocchero, F. *et al.* The hot pick-up technique for batch assembly of van der Waals heterostructures. *Nat. Commun.* **7,** (2016).

24. Purdie, D. G. *et al.* Cleaning interfaces in layered materials heterostructures. *arXiv Prepr. arXiv1803.00912* (2018).

25. Banszerus, L. *et al.* Ultrahigh-mobility graphene devices from chemical vapor deposition on reusable copper. *Sci. Adv.* **1,** e1500222 (2015).

26. Banszerus, L. *et al.* Ballistic transport exceeding 28 μm in CVD grown graphene. *Nano Lett.* **16,** 1387–1391 (2016).

27. Weatherup, R. S. *et al.* Long-term passivation of strongly interacting metals with single-layer graphene. *J. Am. Chem. Soc.* **137,** 14358–14366 (2015).

28. Piquemal-Banci, M. *et al.* Insulator-to-metallic spin-filtering in 2D-magnetic tunnel junctions based on hexagonal boron nitride. *ACS Nano* (2018).

29. Kidambi, P. R. *et al.* Observing graphene grow: catalyst–graphene interactions during scalable graphene growth on polycrystalline copper. *Nano Lett.* **13,** 4769–4778 (2013).

30. Preobrajenski, A. B., Vinogradov, A. S. & Mårtensson, N. Monolayer of h-BN

chemisorbed on Cu(1 1 1) and Ni(1 1 1): The role of the transition metal 3d states. *Surf. Sci.* **582,** 21–30 (2005).

31. Alcock, C. B., Itkin, V. P. & Horrigan, M. K. Properties of the elements and inorganic compounds, vapour pressure of metallic elements. *Can. Met. Q* **23,** 309–313 (1984).

32. Hayne, W. *Melting, boiling, triple, and critical points of the elements. CRC handbook of chemistry and physics, 97th edition.* (CRC Press, 2016).

33. Paffett, M. T., Simonson, R. J., Papin, P. & Paine, R. T. Borazine adsorption and decomposition at Pt(111) and Ru(001) surfaces. *Surf. Sci.* **232,** 286–296 (1990).

34. Preobrajenski, A. B., Nesterov, M. A., Ng, M. L., Vinogradov, A. S. & Mårtensson, N. Monolayer h-BN on lattice-mismatched metal surfaces: On the formation of the nanomesh. *Chem. Phys. Lett.* **446,** 119–123 (2007).

35. Gao, Y. *et al.* Repeated and controlled growth of monolayer, bilayer and few-layer hexagonal boron nitride on Pt foils. *ACS Nano* **7,** 5199–5206 (2013).

36. Park, J.-H. *et al.* Large-area monolayer hexagonal boron nitride on Pt foil. *ACS Nano* **8,** 8520–8528 (2014).

37. Babenko, V. *et al.* Rapid epitaxy-free graphene synthesis on silicidated polycrystalline platinum. *Nat. Commun.* **6,** 7536 (2015).

38. Kilday, M. V, Johnson, W. H. & Pros, E. J. Heat of combustion of borazine. *J. Res. Natl. Bur. Stand. (1934).* **65,** 101–104 (1961).

39. Babenko, V. *et al.* Time dependent decomposition of ammonia borane for the controlled production of 2D hexagonal boron nitride. *Sci. Rep.* **7,** 1–12 (2017).

40. Trehan, R., Lifshitz, Y. & Rabalais, J. W. Auger and x-ray electron spectroscopy studies

of h-BN, c-BN, and $N_2^+$ ion irradiation of boron and boron nitride. *J. Vac. Sci. Technol. A Vacuum, Surfaces, Film.* **8,** 4026–4032 (1990).

41. Ma, T. *et al.* Repeated growth-etching-regrowth for large-area defect-free single-crystal graphene by chemical vapor deposition. *ACS Nano* **8,** 12806–12813 (2014).

42. Li, L. H., Cervenka, J., Watanabe, K., Taniguchi, T. & Chen, Y. Strong oxidation resistance of atomically thin boron nitride nanosheets. *ACS Nano* **8,** 1457–1462 (2014).

43. Liu, Z. *et al.* Ultrathin high-temperature oxidation-resistant coatings of hexagonal boron nitride. *Nat. Commun.* **4,** (2013).

44. Predel, B. B-Pt (Boron-Platinum): Datasheet from Landolt-Börnstein - Group IV Physical Chemistry · Volume 5B: 'B-Ba – C-Zr' in SpringerMaterials (http://dx.doi.org/10.1007/10040476_380). doi:10.1007/10040476_380

45. Weatherup, R. S., Dlubak, B. & Hofmann, S. Kinetic control of catalytic CVD for high-quality graphene at low temperatures. *ACS Nano* **6,** 9996–10003 (2012).

46. Tanuma, S., Powell, C. J. & Penn, D. R. Calculations of electron inelastic mean free paths. IX. Data for 41 elemental solids over the 50 eV to 30 keV range. *Surf. Interface Anal.* **43,** 689–713 (2011).

47. Weatherup, R. S. *et al.* In situ graphene growth dynamics on polycrystalline catalyst foils. *Nano Lett.* **16,** 6196–6206 (2016).

48. Brown, L. *et al.* Polycrystalline graphene with single crystalline electronic structure. *Nano Lett.* **14,** 5706–5711 (2014).

49. Nguyen, V. L. *et al.* Seamless stitching of graphene domains on polished copper (111) foil. *Adv. Mater.* **27,** 1376–1382 (2015).

50. Thompson, C. V. Grain growth in thin films. *Annu. Rev. Mater. Sci.* **20,** 245–268 (1990).

51. Lee, D. S. *et al.* Preferred orientation controlled giant grain growth of platinum thin films on SiO 2/Si substrates. *Japanese J. Appl. Physics, Part 2 Lett.* **40,** 10–13 (2001).

52. Gladman, T. *Grain size control*. (OCP science Philadelphia, Pa/USA, 2004).

53. Humphreys, F. J. & Hatherly, M. *Recrystallization and related annealing phenomena*. (Elsevier, 2012).

54. Beck, G. & Bachmann, C. Oxygen removal at grain boundaries in platinum films on YSZ. *Solid State Ionics* **262,** 508–511 (2014).

55. Kim, M. H. *et al.* Changes in preferred orientation of Pt thin films deposited by DC magnetron sputtering using Ar/O2 gas mixtures. *J. Mater. Res.* **14,** 1255–1260 (1999).

56. Preußner, J., Fleischmann, E., Völkl, R. & Glatzel, U. Enrichment of boron at grain boundaries of platinum-based alloys determined by electron energy loss spectroscopy in a transmission electron microscope. *Int. J. Mater. Res.* **101,** 577–579 (2010).

57. Davis, J. R. *Copper and copper alloys*. (ASM international, 2001).

58. Müller, F., Stöwe, K. & Sachdev, H. Symmetry versus commensurability: epitaxial growth of hexagonal boron nitride on Pt(111) from B-Trichloroborazine (ClBNH)$_3$. *Chem. Mater.* **17,** 3464–3467 (2005).

59. Yao, Y. *et al.* Graphene cover-promoted metal-catalyzed reactions. *Proc. Natl. Acad. Sci.* **111,** 17023–17028 (2014).

60. Ng, M. L. *et al.* Reversible modification of the structural and electronic properties of a boron nitride monolayer by CO intercalation. *ChemPhysChem* **16,** 923–927 (2015).

61. Blume, R. *et al.* The influence of intercalated oxygen on the properties of graphene on

polycrystalline Cu under various environmental conditions. *Phys. Chem. Chem. Phys.* **16,** 25989–26003 (2014).

62. Gorbachev, R. V *et al.* Hunting for monolayer boron nitride: optical and Raman signatures. *Small* **7,** 465–468 (2011).

63. Cai, Q. *et al.* Raman signature and phonon dispersion of atomically thin boron nitride. *Nanoscale* **9,** 3059–3067 (2017).

64. Schué, L., Stenger, I., Fossard, F., Loiseau, A. & Barjon, J. Characterization methods dedicated to nanometer-thick hBN layers. *2D Mater.* **4,** 015028 (2017).

65. Ferrari, A. C. Raman spectroscopy of graphene and graphite: Disorder, electron–phonon coupling, doping and nonadiabatic effects. *Solid State Commun.* **143,** 47–57 (2007).

66. Dean, C. R. *et al.* Boron nitride substrates for high-quality graphene electronics. *Nat. Nanotechnol.* **5,** 722–726 (2010).

67. Lee, J. E., Ahn, G., Shim, J., Lee, Y. S. & Ryu, S. Optical separation of mechanical strain from charge doping in graphene. *Nat. Commun.* **3,** 1024–1028 (2012).

68. Neumann, C. *et al.* Raman spectroscopy as probe of nanometre-scale strain variations in graphene. *Nat. Commun.* **6,** 8429 (2015).

69. Deemyad, S. & Silvera, I. F. Temperature dependence of the emissivity of platinum in the IR. *Rev. Sci. Instrum.* **79,** 10–12 (2008).

70. Bluhm, H. *et al.* In situ x-ray photoelectron spectroscopy studies of gas-solid interfaces at near-ambient conditions. *Mrs Bull.* **32,** 1022–1030 (2007).

71. Zwick, A. & Carles, R. Multiple-order Raman scattering in crystalline and amorphous silicon. *Phys. Rev. B* **48,** 6024–6032 (1993).

## Acknowledgements

We acknowledge funding from the ERC (InsituNANO, grant 279342) and EPSRC (EP/K016636/1). R.W. acknowledges EPSRC Doctoral Training Award (EP/M506485/1). F.M. acknowledges funding from EPSRC Grant No. EP/P00945X/1. R.S.W. acknowledges funding from the European Union's Horizon 2020 research and innovation programme through a EU Marie Skłodowska-Curie Individual Fellowship (Global) under grant ARTIST (no. 656870). We wish to thank M.A.L. for proofreading.

## Corresponding Author

*Stephan Hofmann: Email sh315@cam.ac.uk

# Figures

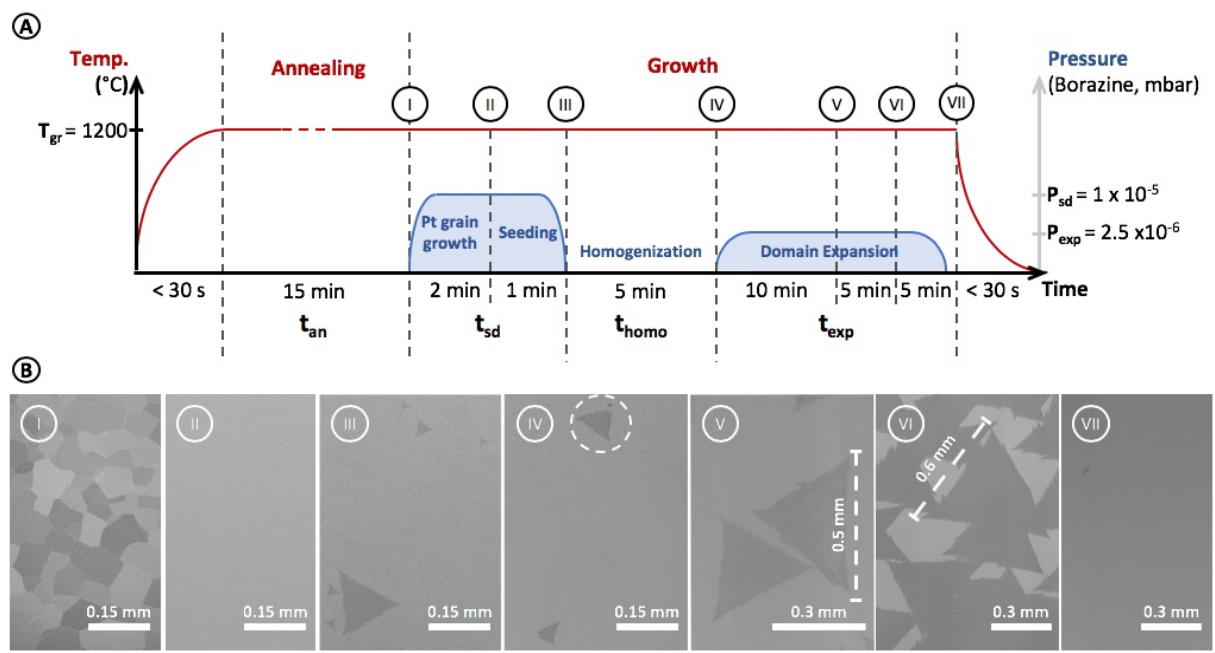

**Figure 1 (a)** Process flow diagram of SSG. The growth temperature is $T_{gr}$=1200 °C. Precursor pressure during seeding is $P_{sd}$=1x10$^{-5}$ mbar and $P_{exp}$=2.5x10$^{-6}$ mbar during domain expansion **(b)** SEM images of h-BN on Pt at different stages of growth. Growth was stopped at the respective stages, by removing the precursor and turning off the laser heating. In image IV, after annealing of nuclei, damage to existing domains is clearly visible (dotted outline).

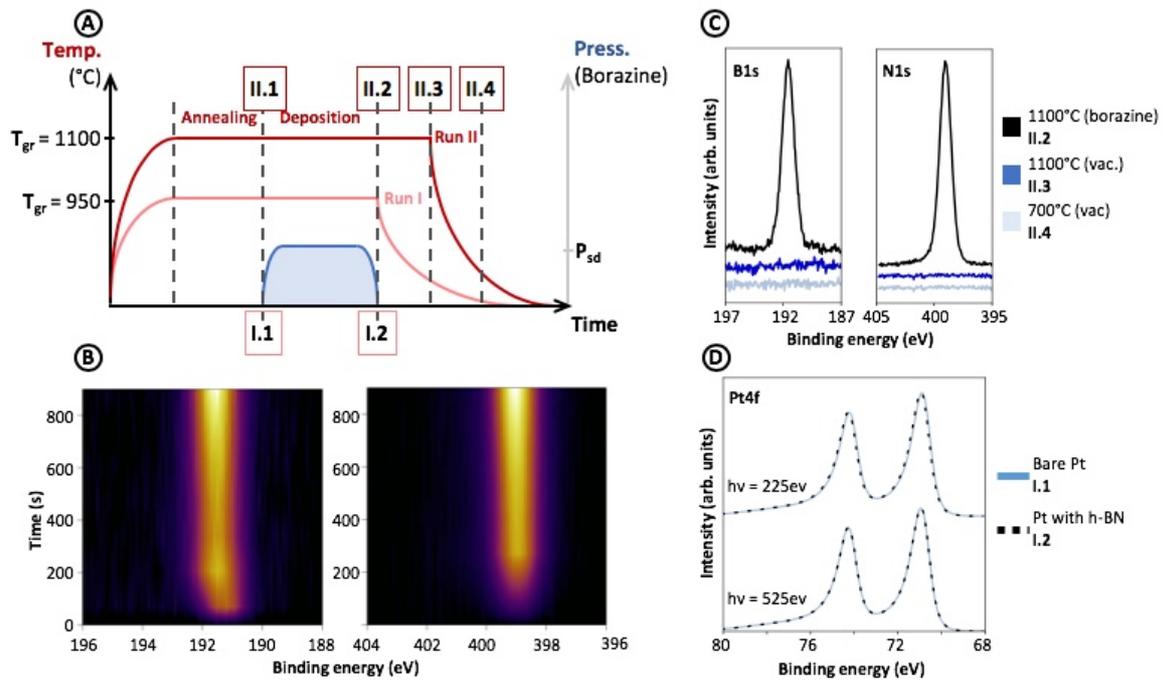

**Figure 2** In-situ XPS measurements during h-BN growth on Pt foil. Details on the conditions are given in the experimental section. **(a)** Schematic process flow diagram highlighting at which point of the process the spectra shown in (c), (d), and (e) were taken. **(b)** Evolution of the B 1s and N 1s XP core level with borazine ($3\times10^{-4}$ mbar) exposure time (spectra taken between II.1-II.2) for a Pt foil at 1100 °C **(c)** B 1s and N 1s XP spectra taken at an excitation energy of hv = 620 eV for $T_{gr}$ = 1100 °C with precursor present, $T_{gr}$ = 1100°C in vacuum and RT in vacuum. The peak positions of B1s/N1s are 191.6eV/399.0eV. Shortly after removing borazine, the B/N peaks disappear and do not reappear during cooling. **(d)** Depth resolved Pt 4f XP spectra taken for Pt covered with h-BN and bare Pt at $T_{gr}$ = 950°C. No difference in peak positions and/or additional peaks are visible, confirming the absence of potential Pt compounds.

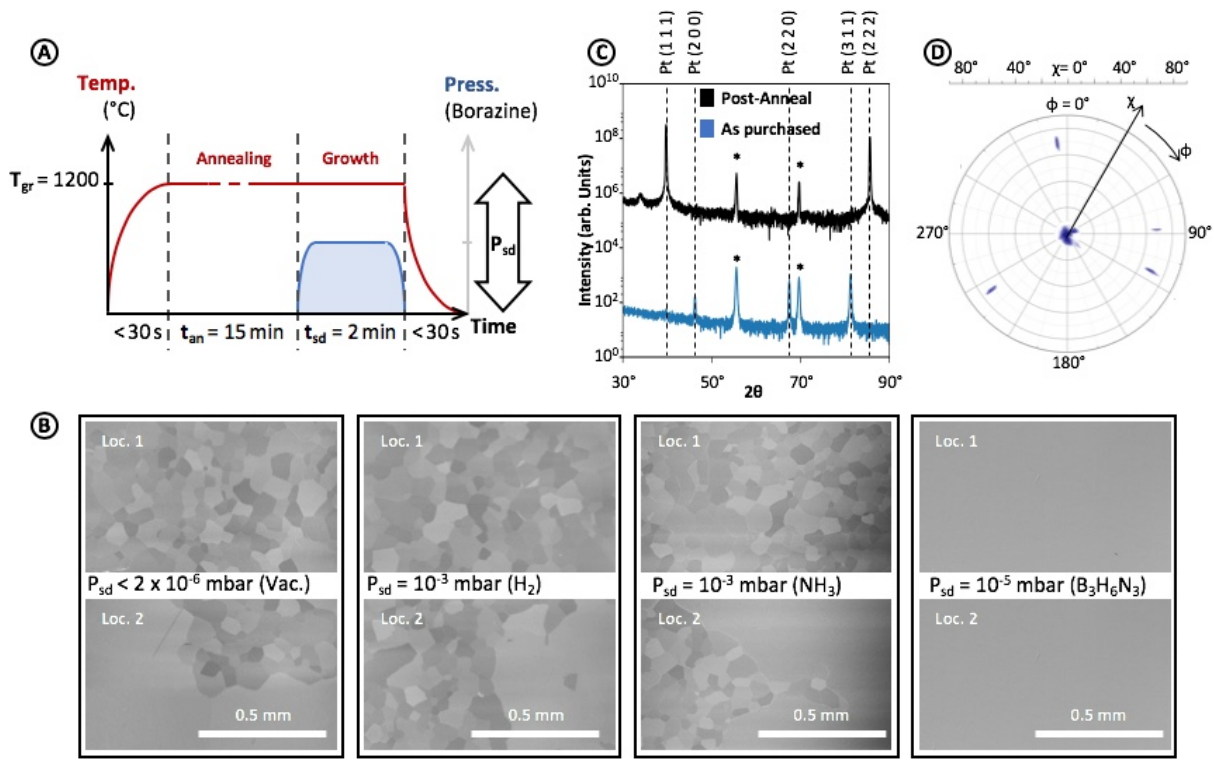

**Figure 3** One-step process at identical $T_{gr}$=1200°C for 2 min, but with different gas environments. (**a**) Process flow diagram for recrystallization baseline experiment. All parameters were held constant, only type of gas and the pressure was varied. (**b**) SEM images of Pt foils after annealing. For each experimental condition, SEM images of two locations on the same substrate were provided to highlight the differences in texture. When annealing in vacuum, $H_2$, or $NH_3$, although certain polycrystalline regions remain (top images), it is apparent that the growth of large single crystal regions has occurred (bright region in bottom images). In contrast, the sample treated with borazine shows no polycrystalline regions. (**c**) XRD measurement of Pt foils as purchased and after recrystallization in borazine ($P_{sc}$ = $10^{-5}$ mbar, $T_{gr}$ = 1200 °C, $t_{an}$ = 15 min and. $t_{sd}$ = 2 min). The spectra have been offset for better visibility (Post–Anneal spectrum was multiplied with a factor of $10^4$). All Pt related peaks are marked. The peaks marked with * originate from the Ta susceptor on which the Pt foil is mounted. While there are multiple orientations for the pristine foil, the dominating orientation is (1 1 1) after annealing. (**d**) Texture map of the Pt (1 1 1) reflection at 2θ = 39.73°. The Pt foil behaves like a single crystal, with one pole in the symmetric position (χ ~ 0°) and 3 poles at χ ~ 70° and φ = 120° apart from each other. A minor pole is visible at χ ~ 70° and φ = 90°, which indicates a minority of differently oriented grains.

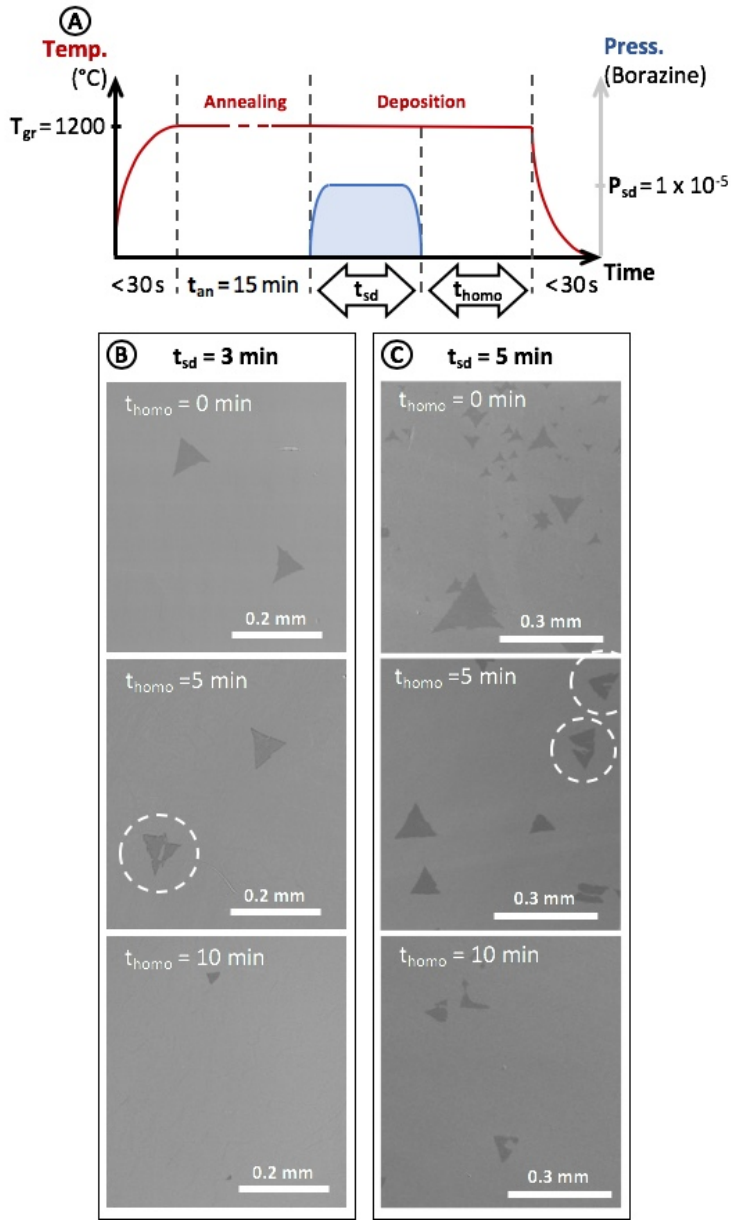

**Figure 4 (a)** Schematic process flow diagram for SG experiment to highlight the effect of seeding and homogenization. **(b)-(c)** SEM images of growth result. Growth was stopped at the respective stages, by removing the precursor and turning off the laser heating All parameters were kept constant, except of $t_{sd}$, which is varied between the series [3min in (e) and in 5 min (f)] and $t_{homo}$, which is varied within each of the series.

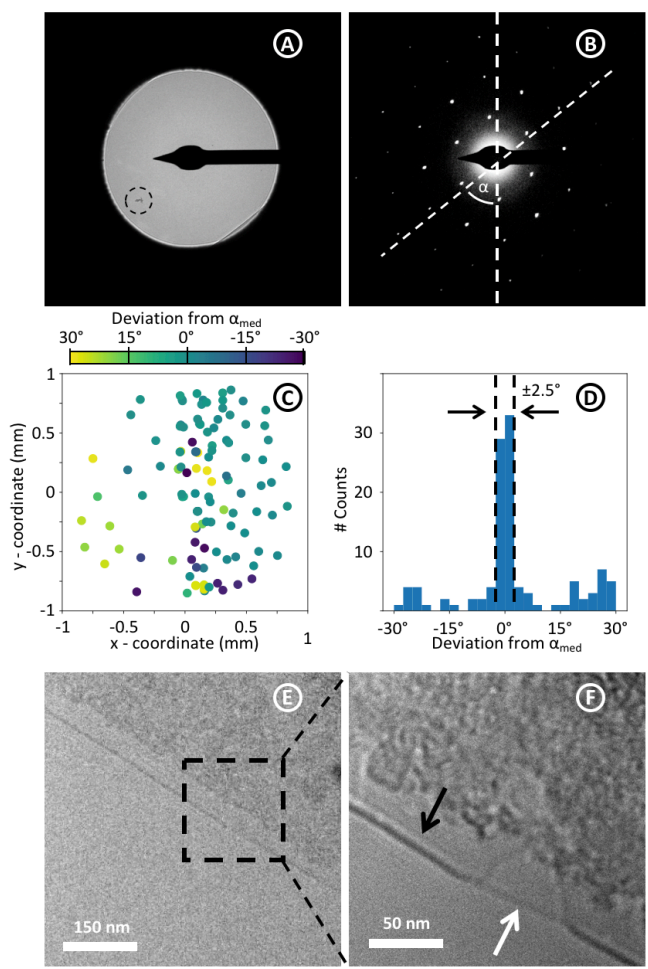

**Figure 5 (a)** BF-TEM images of h-BN. The dotted circle marks a particle on the suspended h-BN, which is otherwise indiscernable. **(b)** DF-TEM image corresponding to (a). α is defined as the angle between the vertical and the closest first order diffraction spot in clockwise direction. **(c)** Scatter map of the rotational deviation from $α_{med}$ (median value of α). **(d)** Distribution of orientation as deviation from $α_{med}$. **(e) & (f)** High magnification BF-TEM image of the edge of the h-BN film. Only one fringe is found (black arrow), which confirms the monolayer nature of BN. A small dent can be seen (white arrow) caused by folding of the layer. The contrast between single layer edge and folded edge is clearly visible.

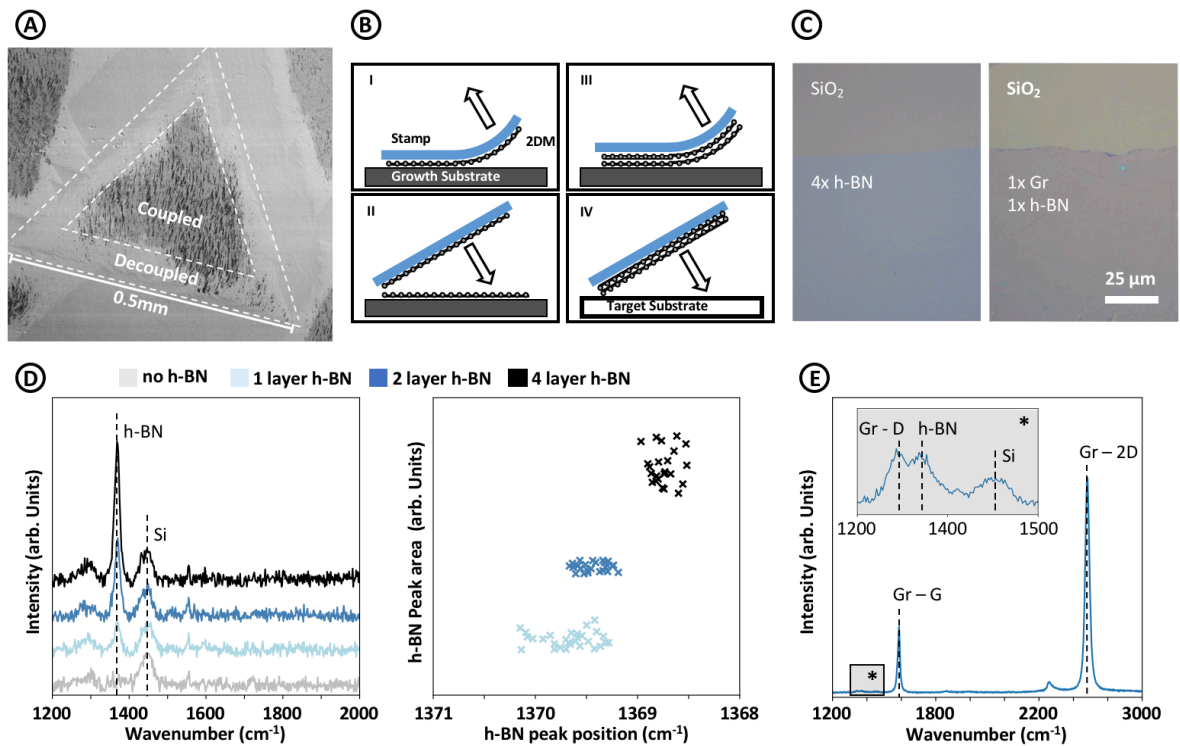

**Figure 6 (a)** SEM image of h-BN on Pt. The image is taken 5 hours after removing the sample from the reactor. The dotted lines mark the outer edge of the domains and the limit between the coupled (black) and decoupled (white) regions. **(b)** Process flow diagram of exfoliation based transfer. The PVA stamp is drop-cast onto the as-grown h-BN, which can then be peeled off and used for sequential exfoliation. After transfer onto the target substrate the stamp is dissolved in water **(c)** Optical images after transfer onto $SiO_2$ of 4-layer h-BN and h-BN/Gr (Gr in contact with $SiO_2$) **(d)** The left graph shows the Raman spectrum of h-BN after transfer onto $SiO_2$ depending on the layer number. The spectra have been offset for better visibility. Si marks the 3rd order silicon peak at ~1450cm$^{-1}$ [71]. The peak at ~1370cm$^{-1}$ corresponds to h-BN [63]. The right-hand plot presents the peak area after fitting with a Lorentzian curve, against the peak position for multiple measurements of different numbers of h-BN layers. For better visibility, only points between the 1st and 3rd quartile are shown for each sample. The median of the peak position is 1369.7 cm$^{-1}$, 1369.4 cm$^{-1}$, 1368.7 cm$^{-1}$ and the median normalized peak area is 1, 2.24 and 4.03 for monolayer, bilayer and 4-layer h-BN. **(e)** Raman spectrum of h-BN/Gr stack after transfer (Gr in contact with $SiO_2$). Inset shows magnified region to highlight the h-BN peak

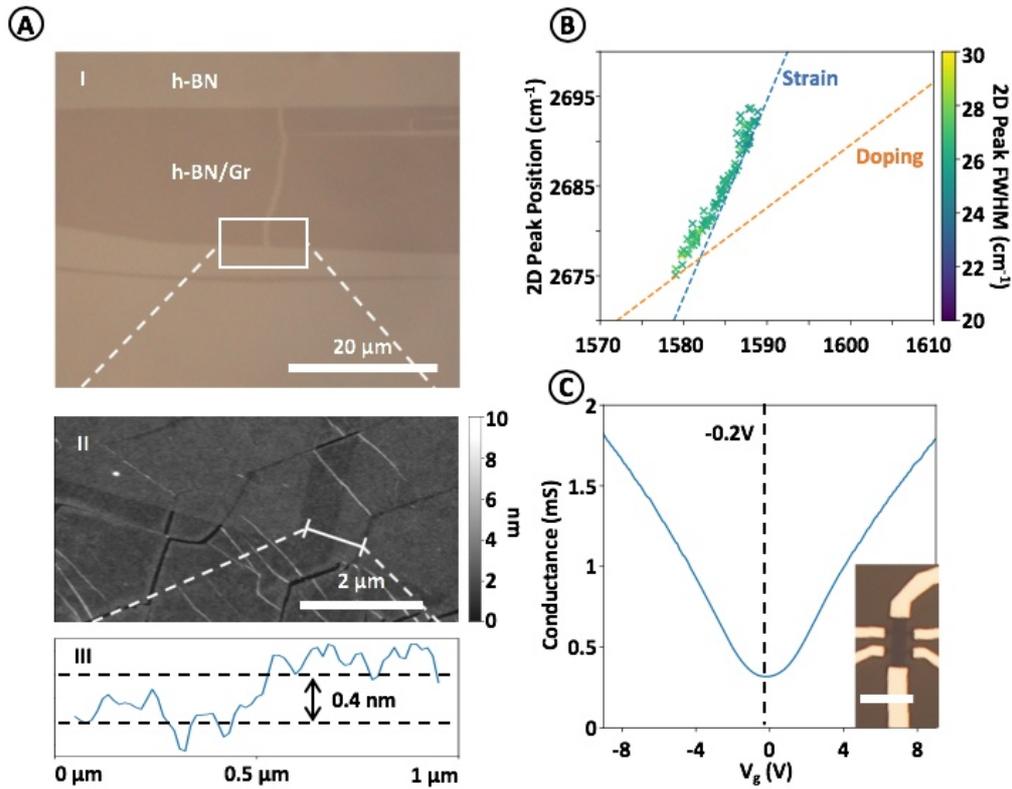

**Figure 7 (a.I)** Optical image of monolayer CVD h-BN on exfoliated Gr transferred onto a 90 nm SiO$_2$/Si wafer (Gr between h-BN and SiO$_2$). h-BN is not discernible as it uniformly covers the sample. **(a.II)** Peak force atomic force microscope (PF-AFM, details in experimental section) of area marked in (a.I). **(a.III)** Profile of line marked in (a.II). Step height of transition from h-BN only to h-BN/Gr region is about ~0.4 nm, as expected for single layer graphene, indicating a clean interface. **(b)** Peak position of the G- and 2D-peak of Gr measured by Raman spectroscopy. Colour of cross relates to FWHM of the 2D peak (See associated colour bar). The dotted blue line is the strain axis (slope 2.2), dotted orange line the doping axis (slope 0.7) and the charge neutrality point is (1582 cm$^{-1}$, 2677 cm$^{-1}$).[67] **(c)** Transfer curve obtained via 4-terminal measurement, details given in experimental methods section. The position of the Dirac point is marked. An optical image of the Hall bar measured is shown in the inset. The scale bar indicates 10 μm.

# Supplementary Information:

# A peeling approach for integrated manufacturing of large mono-layer h-BN crystals

Ruizhi Wang[†], David Purdie[†,‡], Ye Fan[†], Fabien Massabuau[§], Philipp Braeuninger-Weimer[†], Oliver J. Burton[†], Raoul Blume[∥], Robert Schloegl[⊥], Antonio Lombardo[†,‡], Robert S. Weatherup[∇,o], Stephan Hofmann*,[†]

[†]Department of Engineering, University of Cambridge, 9 JJ Thomson Avenue, Cambridge CB3 0FA, United Kingdom

[‡]Cambridge Graphene Centre, University of Cambridge, 9 JJ Thomson Avenue, Cambridge CB3 0FA, United Kingdom

[§]Department of Materials Science and Metallurgy, University of Cambridge, 27 Charles Babbage Road, Cambridge CB3 0FA, United Kingdom

[∥]Helmholtz-Zentrum Berlin für Materialen und Energie, D-12489 Berlin, Germany

[⊥]Fritz Haber Institute, D-14195 Berlin-Dahlem, Germany

[∇] School of Chemistry, University of Manchester, Oxford Road, Manchester M13 9PL, UK

[o] University of Manchester Harwell Campus, Diamond Light Source, Didcot, Oxfordshire, OX11 0DE, UK

# Sample Mounting

During all our growth experiments the Pt foil was placed on a Ta foil, which acted as the susceptor for the laser. It spread the thermal energy and thus allows us to achieve homogeneous heating of the Pt sample, which is bigger than the laser spot size of about 5 mm x 5 mm. The result of heating the sample directly is shown in Fig. S1. In this case, the central region of the foil is significantly hotter (see Fig. S1b) compared to the edge of the sample (see Fig. S1a). This results in inhomogeneous growth, which motivated the application of a susceptor for uniform heating.

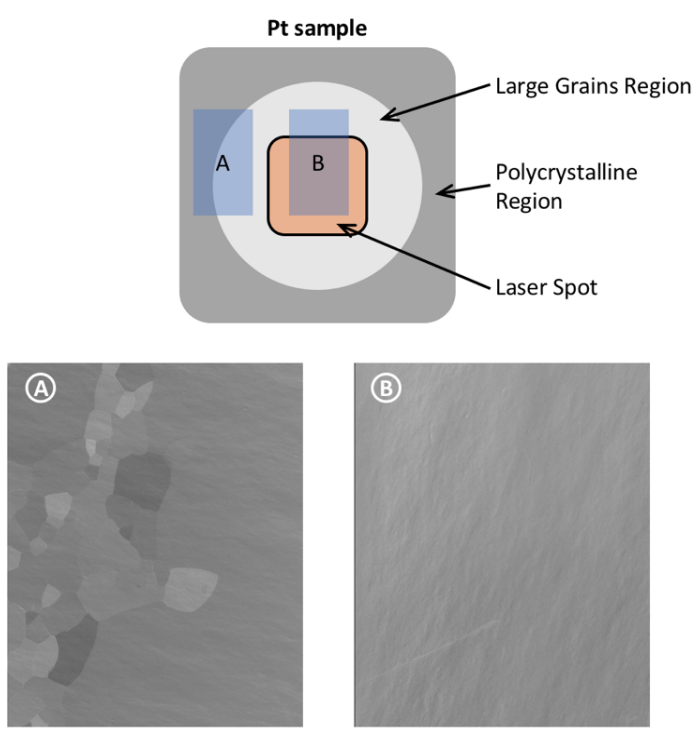

**Figure S1** Direct heating of Pt foil. Sample was annealed at 1250 °C for 1h in 0.5 mbar $H_2$. SEM mages (a) and (b) are taken from different location of the foil as shown schematically in the diagram.

# In-situ XPS Analysis

## Analysis of BN compound

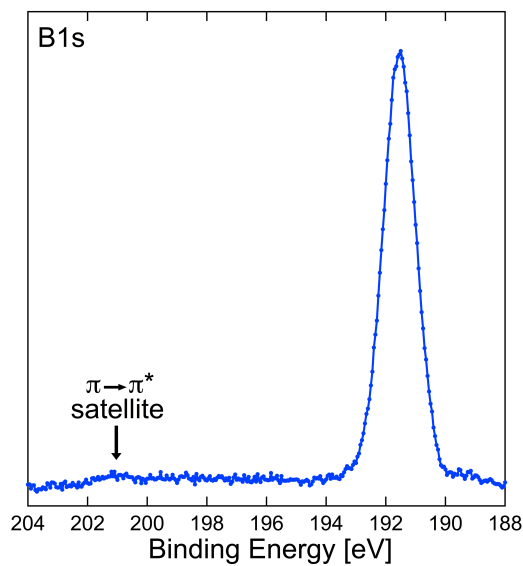

**Figure S2** Summed B 1s XPS core level spectrum consisting of 7 spectra acquired consecutively during borazine ($3\times10^{-4}$ mbar) exposure for a Pt foil at 1100 °C, between 200-1100 s after borazine introduction. The improved signal to noise ratio allows the π→π* plasmon shake up satellite to be more clearly resolved.

## Shift of XPS peaks associated with h-BN

Although the B1s and N1s peaks remain at a relatively constant BE for the majority of the borazine exposure, the peak in the B1s spectrum in Fig. 2 from the main article shows an initial increase in BE (from ~191.2 eV to ~191.5 eV). Throughout our XPS measurements of h-BN growth on Pt, we also observe that the B1s and N1s peak positions change as the photon energy is varied, which for the synchrotron beamline used for these measurements also corresponds to changing the x-ray flux. The size of the shift generally increases as the x-ray flux is increased, consistent with charging of the h-BN. Given the insulating properties of h-BN, we therefore attribute the shifting B1s peak position at the start of growth to the gradual charging of the h-BN layer as it forms on the Pt surface. To account for this, the B1s and N1s spectra presented here are measured with the same photon energy (and thus flux) to ensure that any charging affects both spectra in a similar way, such that their relative peak positions should not be altered. Indeed, the relative B1s and N1s positions match closely with those previously reported.[1–6] However, it should be noted that the values of peak position should not be taken as absolute. There may also be shifts be related to changes in the interaction between h-BN and the Pt substrate with temperature and gas environment, but these are not readily disentangled from the observed charging effects.

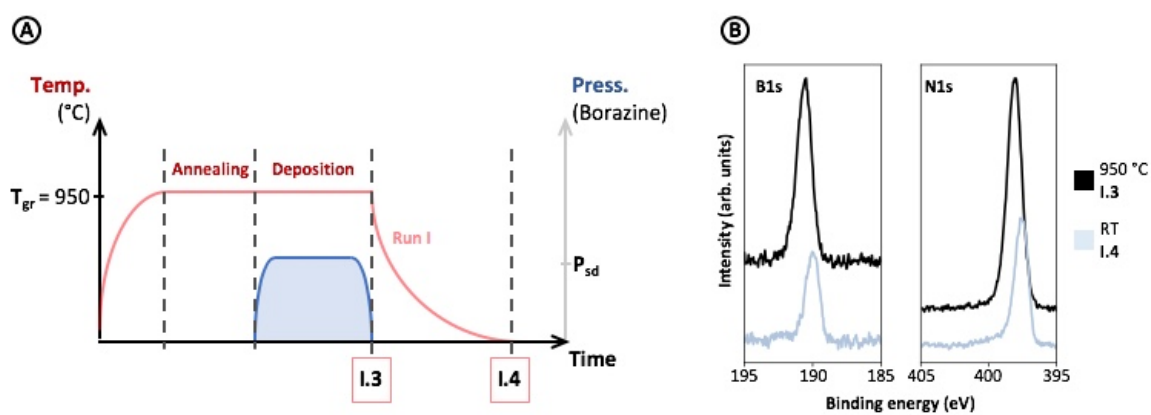

**Figure S3 (a)** Schematic process flow diagram highlighting at which point of the process the spectra shown in (b) were recorded. **(b)** B1s and N1s XP spectra taken at an excitation energy of hν=620eV for $T_{gr}$ = 950 °C and RT. The positions of the peak centres are 190.6eV/398.1eV at $T_{gr}$ and 190.0eV/397.6eV

## h-BN precipitation

In order to check, whether the h-BN is dissolved into the bulk during the homogenization step in SSG, we performed a series of experiments with the goal of growth by precipitation. Samples were grown at high temperature ($T_{gr}$ = 1275 ºC) following the SSG procedure. After SEM analysis the same samples were loaded again into the reactor and annealed at a lower temperature ($T_{pr}$ = 950 ºC) without any precursor. After this step, smaller secondary nuclei are formed in addition to the previous ones. We attribute this to growth by precipitation, due to supersaturation achieved by lowered solubility. These secondary nuclei are not formed during initial SSG, as the samples are quenched.

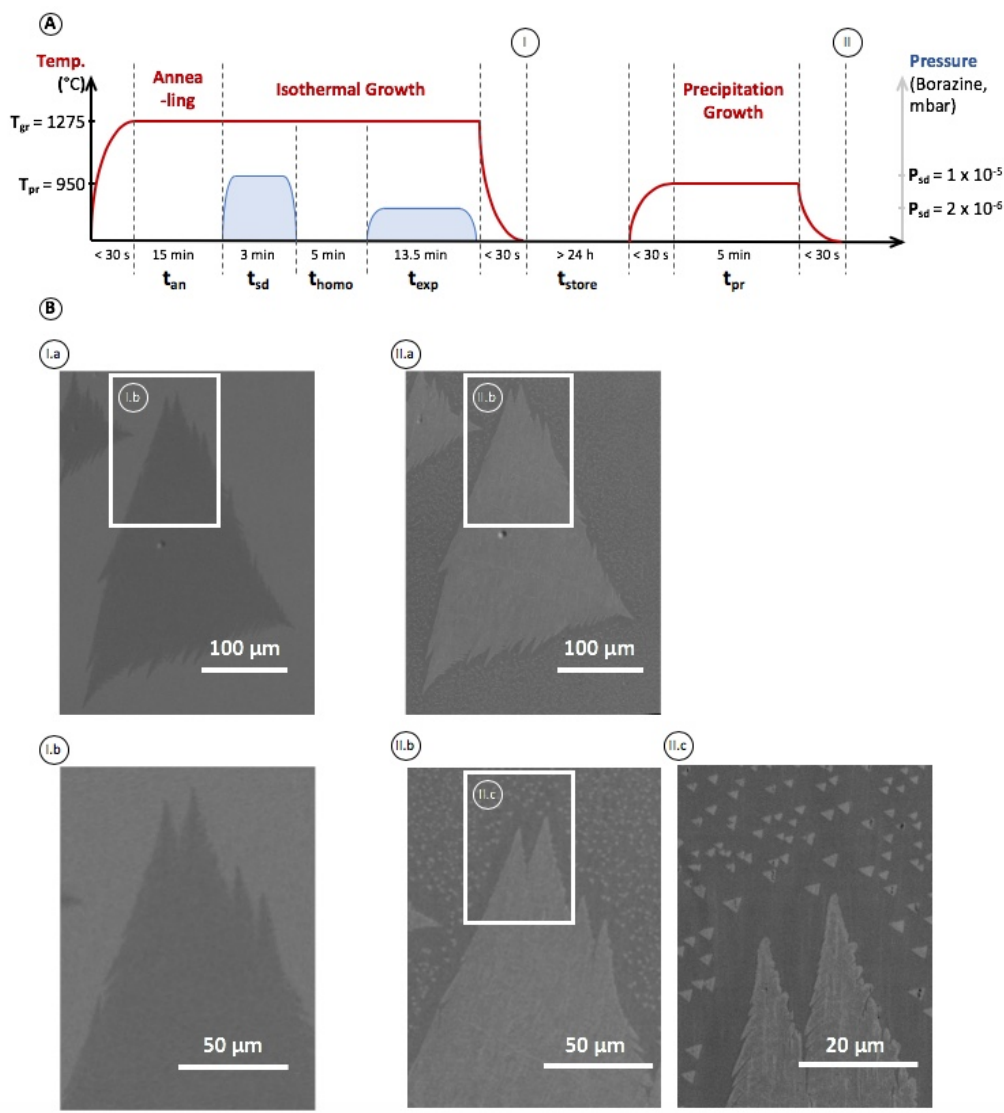

**Figure S4 (a)** Schematic process flow diagram highlighting at which point of the process the spectra shown in (b) were recorded. First the sample was grown using SSG as shown in Fig 1, but at higher temperature. **(b)** (I.a) & (I.b) are SEM images taken of the samples immediately after SSG growth. The sample was quenched post SSG by turning off the heater immediately. (I.b) shows the area marked in (I.a) at higher magnification. (II.a), (II.b) & (II.c) are SEM images taken of the same sample after precipitation growth. The images are taken more than 24h after the process. In this period, the samples were kept in ambient, resulting in the change of contrast associated with intercalation (see Fig 6a in main text and Fig. S13) (II.b) shows the area marked in (II.a) at higher magnification; same for (II.c) and (II.b). Smaller h-BN islands have formed in addition to the previous ones, despite absence of precursor.

The fact that growth can take place by precipitation shows that both boron and nitrogen, even in small amounts, are absorbed into the bulk during SSG. This also explains why in our experiments repeating the cycle of homogenization and domain expansion multiple times (i.e. repeating the steps III – V from Fig. 1) does not result in a noticeable increase in domain size or decrease in nuclei density. The initial motivation for the homogenization step is the reduction of secondary nuclei. It relies on the fact that primary and secondary nuclei differ in size but dissociate at the same rate. However, after a first cycle of homogenization and domain expansion, the given domains will be similar in size. Thus, an additional homogenization step does not result in the desired size-filtering effect. Furthermore, we observe that the dissociation of the domains is considerable slower during homogenization given a previous growth cycle. In previous literature the growth of Gr was tuned by repeatedly growing and dissociating as-grown domains.[7] This was achieved by gas induced etching of the Gr. The attenuation of the dissociation during repeated homogenization in our experiments further corroborates that here it is mainly driven by bulk absorption. Once a significant quantity of species is present in the catalyst, additional absorption in slowed or even halted, which corresponds to our experimental observation.

# Detailed investigation of parameter space for SSG

In order to explore the parameter space of SSG, we performed series of experiments, where individual parameters were changed and the impact on the growth process was analysed. Figs S5 – S8 show the result of these experiments.

## Variation of seeding time ($t_{sd}$)

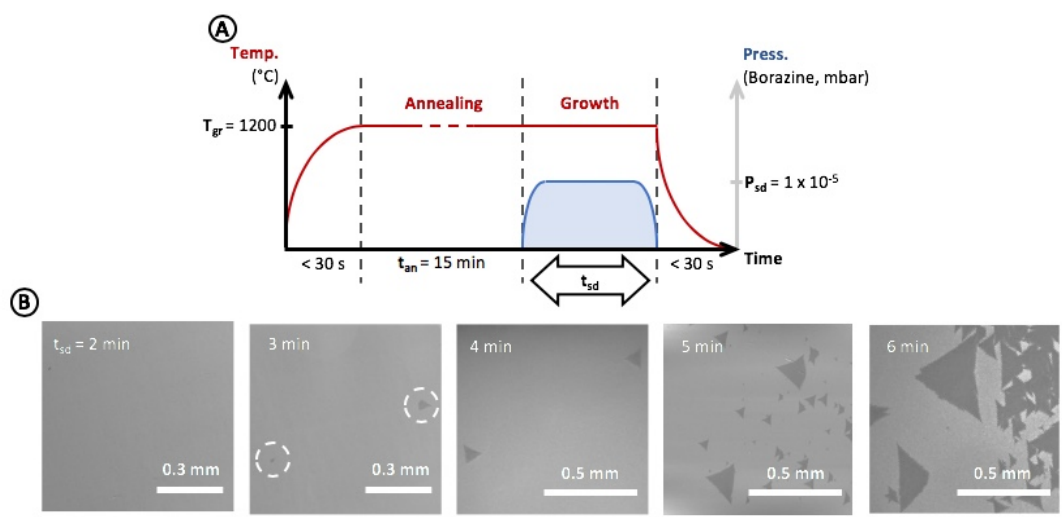

**Figure S5 (a)** Process flow diagram for SG at $T_{gr}$=1200°C and precursor pressure of $P_{sd}$=1x10$^{-5}$mbar, while varying the growth time $t_{sd}$. **(b)** SEM images of the samples after growth. For each SEM image the growth was stopped at the respective time by quenching the sample. After 3 min, the onset of nucleation begins (nuclei marked by white dotted circles). These grow in size, however, at 5 min, secondary nucleation sets in. While there will be very large islands formed eventually (>0.4mm after 6 min), there is a large number of smaller nuclei that prevent their continued growth due to coalescence.

**Variation of seeding pressure ($P_{sd}$ = 2.5 x $10^{-6}$ mbar instead of 1 x $10^{-5}$ mbar)**

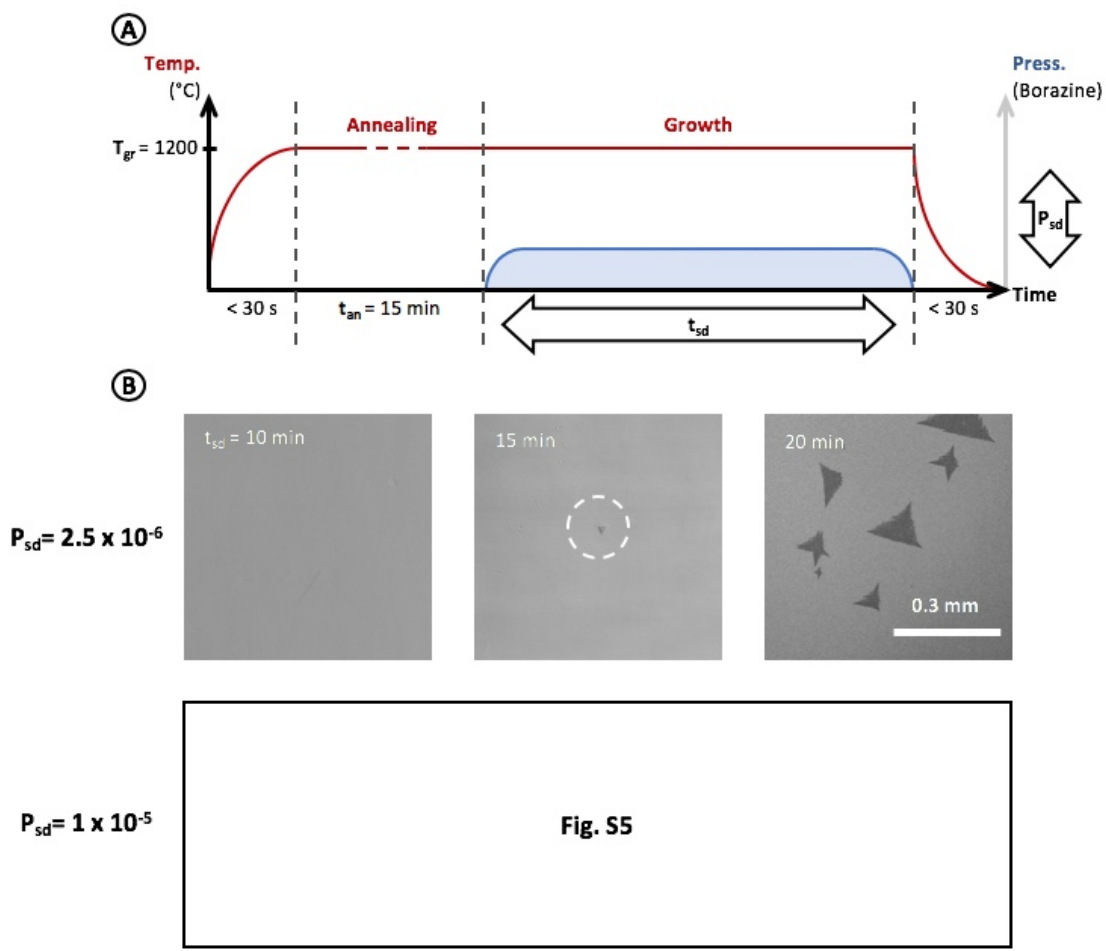

**Figure S6 (a)** SG at $T_{gr}$=1200°C. The precursor pressure of $P_{sd}$ is varied. In order to show the effect on nucleation and growth, the growth was stopped at different times $t_{sd}$. **(b)** Comparison of nucleation between $P_{sd}$ = 2.5 x $10^{-6}$ mbar (shown here) and $P_{sd}$ = 1 x $10^{-6}$ mbar (shown in Fig. S5). Due to the lower precursor pressure, a much higher incubation time is required before the onset of nucleation. After 15 min of precursor exposure, the first nuclei are formed. When extending the exposure time, more nuclei form, as evidenced after 20 min. The scale bar applies to all SEM images.

## Variation of growth temperature ($T_{gr}$)

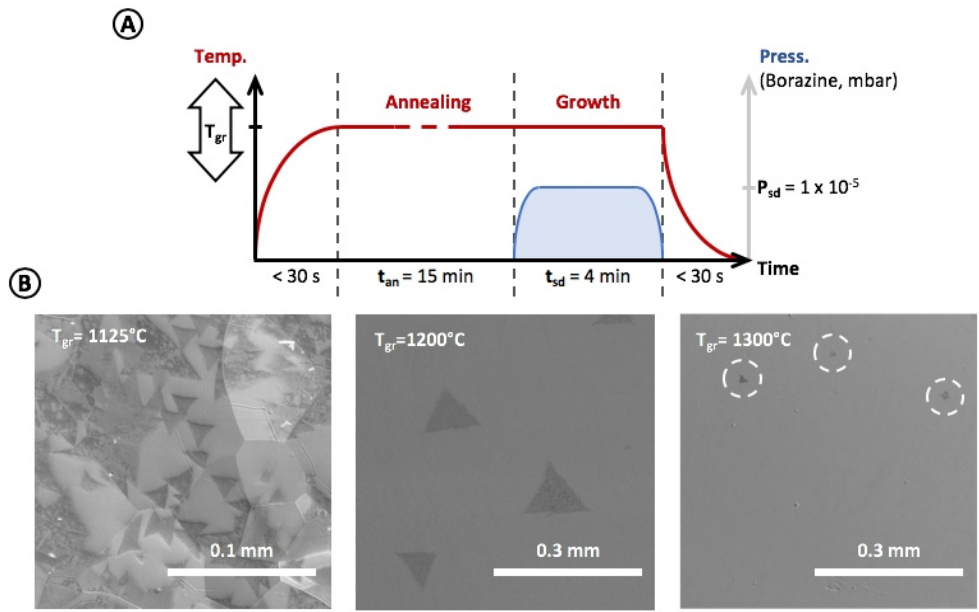

**Figure S7 (a)** SG of h-BN at different temperatures $T_{gr}$, while keeping all other parameters constant. **(b)** For each SEM image the growth was stopped at the respective time by quenching the sample. An increase of temperature leads to reduced nucleation (different scale bars chosen for improved visibility). Increasing the temperature further does not have significant impact on nucleation reduction, but growth is slowed.

**Effect of Pt crystallinity**

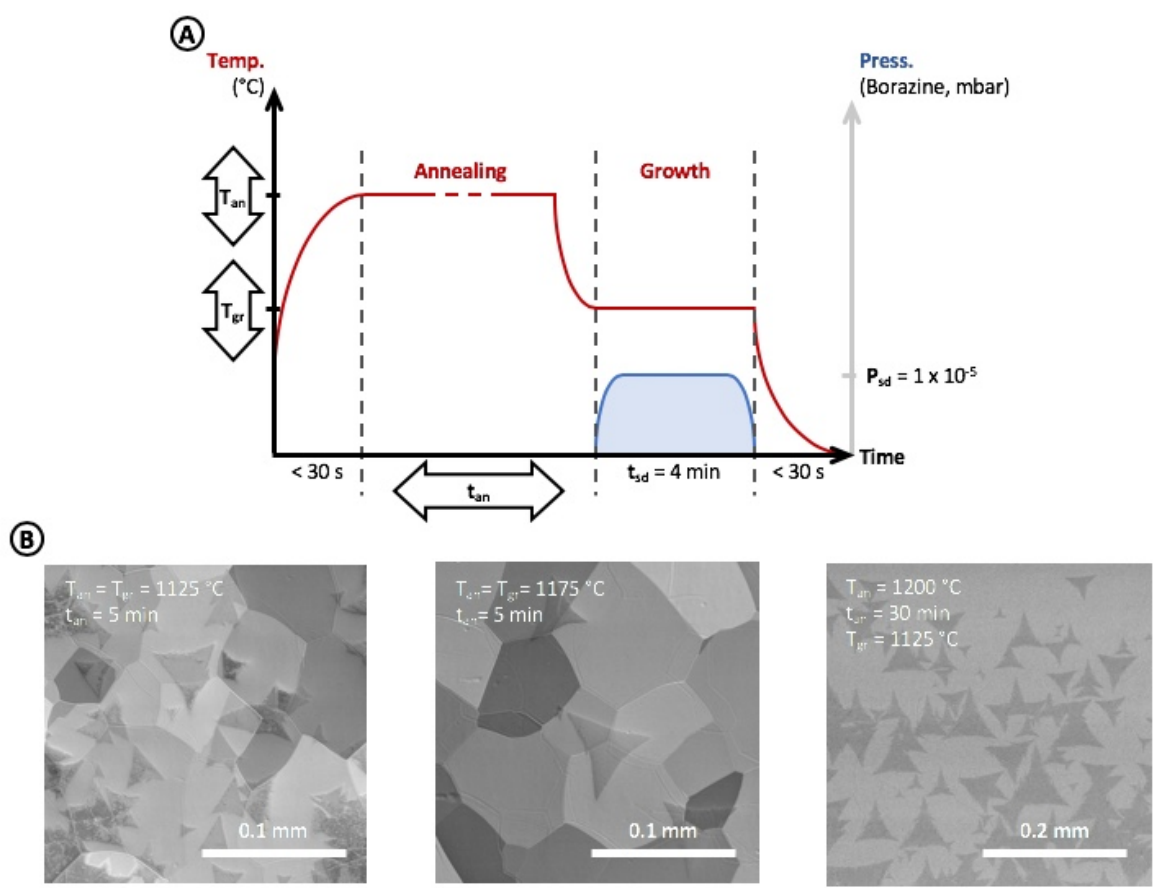

**Figure S8 (a)** SG of h-BN for different growth temperatures $T_{gr}$ and crystallinity of foil. **(b)** For each SEM image the growth was stopped at the respective time by quenching the sample. (Left) Baseline sample grown using standard growth (SG) process at $T_{an} = T_{gr} = 1125$ °C. (Mid) Result of increased growth temperature. All growth parameter chosen identical to left image, but growth temperature was increased to $T_{an} = T_{gr} = 1175$ °C (Right) Change of crystallinity of foil. Identical growth parameter as baseline sample, but foil was pre-annealed at $T_{an} = 1200$ °C for 30 min. The uniform contrast in the background indicates single crystal Pt in field of view.

# Additional sample characterization

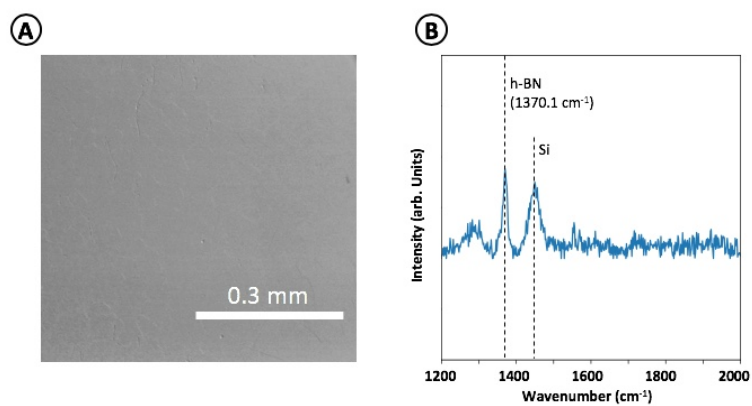

**Figure S9 (a)** Growth of h-BN film on Pt foil. Lack of contrast due to continuous film of h-BN. Image is taken of h-BN grown on Pt foil, which has been used for a previous growth/transfer cycle. During first run a continuous film of h-BN was grown and transferred off using the methods described in the experimental methods section. Before regrowth, the only catalyst preparation step consists of sonicating the sample in water, to remove potential residues of PVA. The second run was performed using a SG process as outlined in Fig. S2 using the conditions: $T_{gr}$ = 1175ºC, $t_{sd}$ = 10min to achieve a continuous layer of h-BN. **(b)** Raman spectrum of sample from image (a). The sample regrown on a previously used Pt foil was transferred using peeling transfer onto SiO$_2$ (300 nm)/Si wafer. The h-BN/ Si peak intensity ratio (~1, compare with Fig. 6) and peak position (1370.1 cm$^{-1}$) indicate monolayer h-BN. The h-BN is of high quality as shown by the FWHM of around 13.5 cm$^{-1}$, which compares well to the FWHM of monolayer bulk exfoliated h-BN.[8]

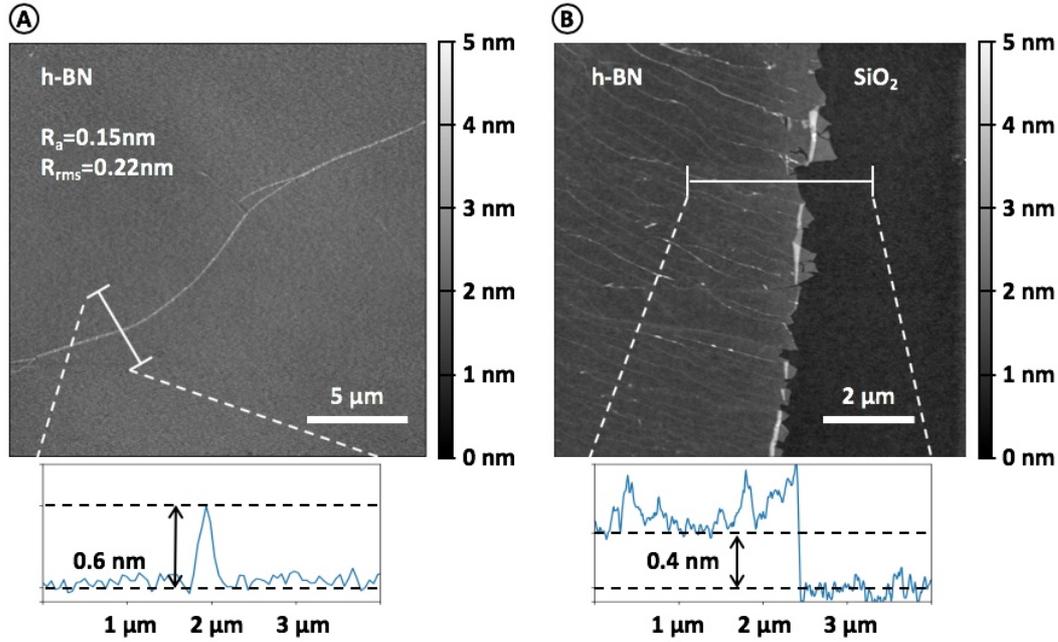

**Figure S 10** Peak force atomic force microscope (PF-AFM, details in experimental section) measurement of h-BN transferred onto $SiO_2$ using direct exfoliation. Prior to the measurements, the sample was heated to 250°C in ambient condition for 5 min. to remove atmospheric adsorbents. **(a)** Height profile of central region of film. The average deviation from mean ($R_a$) and root mean square average deviation from mean ($R_{rms}$) are given for the field of view. Due to the extremely low roughness of the surface, the only visible feature is the wrinkle in the central region. **(b)** Step height measurement at the edge of the film. The height is ~0.4nm, which corresponds to monolayer h-BN. More wrinkles are found on the edge, compared with (A), which is potentially induced by the transfer process.

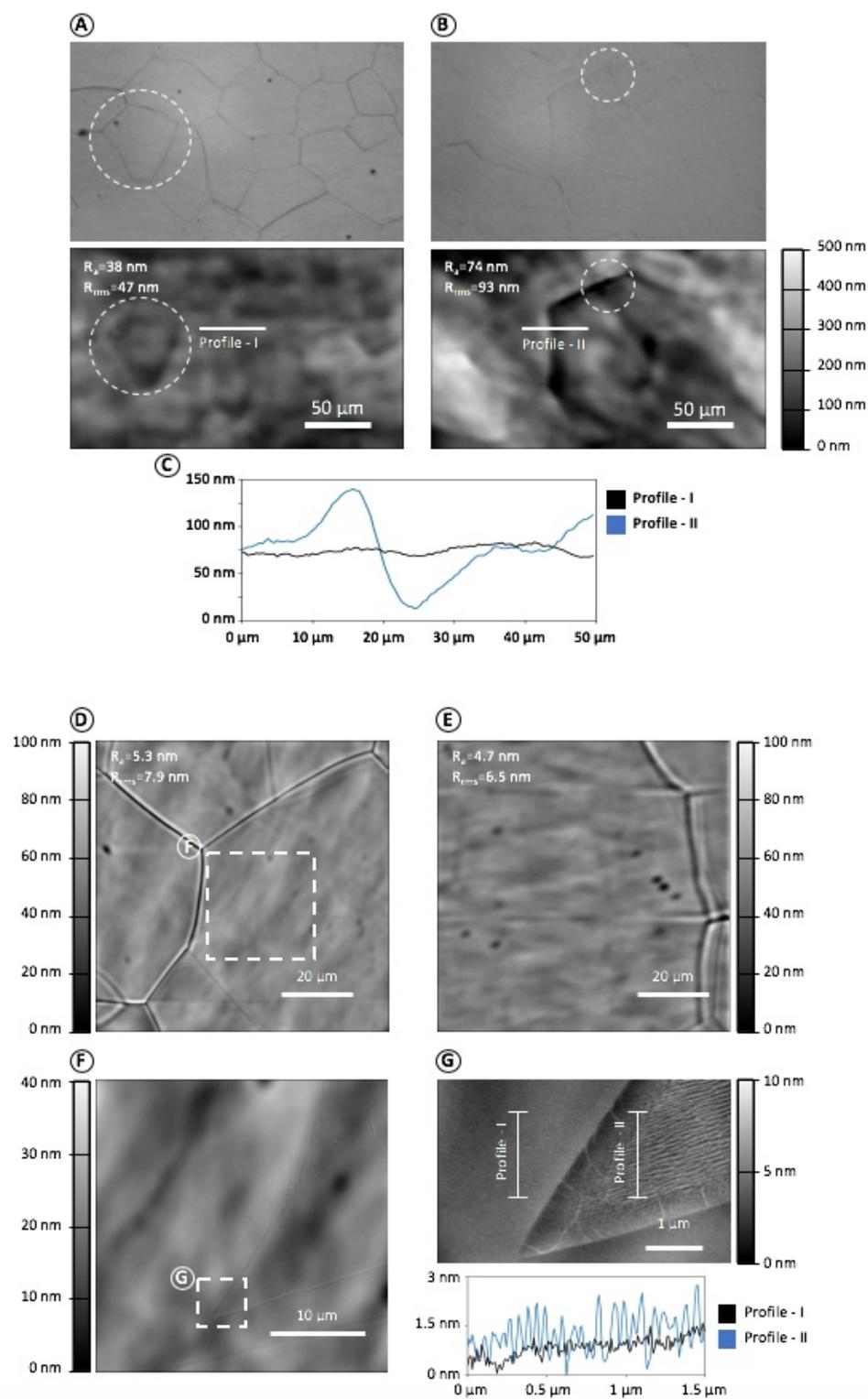

**Figure S 11** White light interferometer (WLI) **(a)-(c)** and PF-AFM **(d)-(g)** measurements of h-BN on Pt after growth for polycrystalline [**(a)**, **(d)**, **(f)** & **(g)**] recrystallized [**(b)** & **(e)**] Pt foil. In **(a)** and **(b)** the surface map by WLI and the corresponding optical image are shown. Common features are highlighted by dotted circle for better visibility. The origin of the apparent difference in roughness is highlighted in **(c)**. The surface profile of the recrystallized foil shows, that the higher roughness originates from a large-scale curvature of the foil. The comparison of PF-AFM images of polycrystalline **(d)** and recrystallized (**e**) foil, shows that on small scales, the roughness is similar. **(f)** & **(g)** are magnified images of **(d)** & **(f)**, corner of an h-BN island. It should be noted that the h-BN is mainly visible due to a change in measured roughness and not step height.

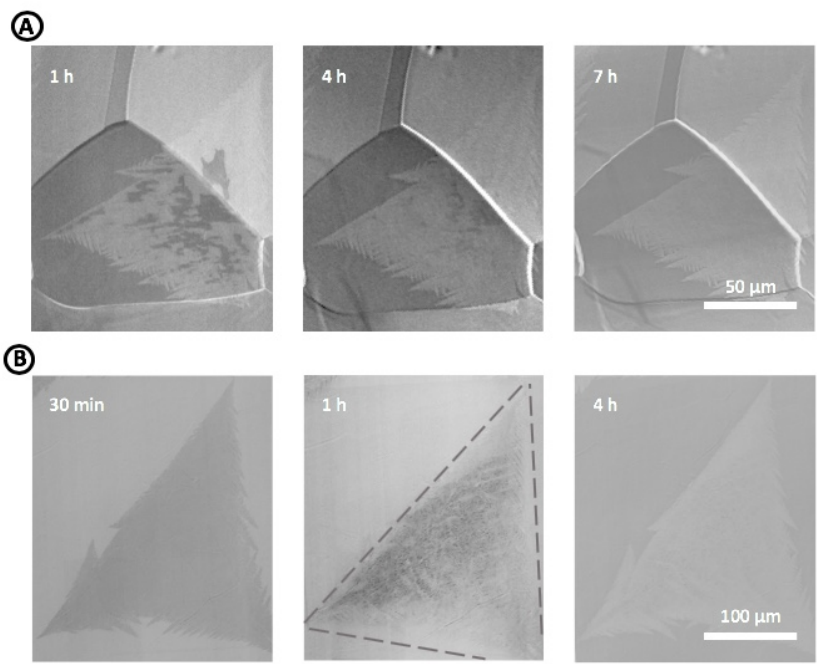

**Figure S 12** SEM images of the evolution of decoupling of h-BN on Pt for two different samples. Both samples were grown using SG. **(a)** Sample grown at $T_{gr}$=1150°C to retain polycrystalline structure of Pt foil. **(b)** Sample grown at $T_{gr}$=1200°C after Pt foil recrystallization

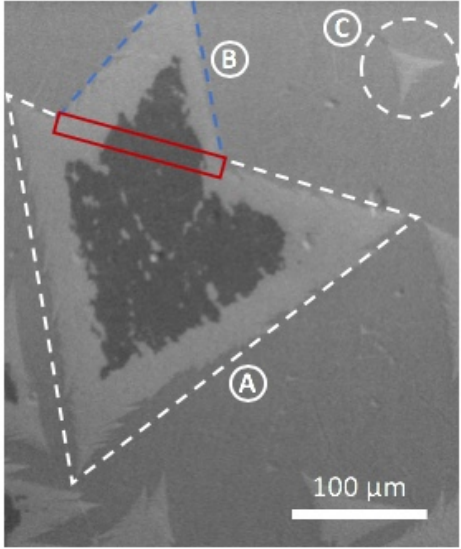

**Figure S 13** SEM image of sample grown at $T_{gr}$=1200°C after Pt foil recrystallization. Image taken about 45 min after growth process. The areas with white dotted lines (A) and blue dotted lines (B) are two h-BN islands in the process of merging. (C) marks another island. Due to the difference in size, (C) has already completely decoupled. Islands (A) & (B) are in the process of decoupling. The red box marks the most likely location for the grain boundary between islands (A) and (B), if any were present. Intercalation is not observed from this location.

# Additional device data

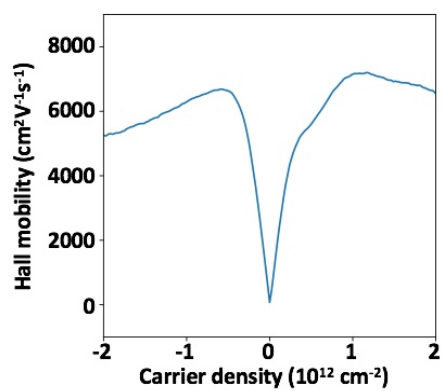

**Figure S 14** Extracted Hall mobility as a function of charge carrier density for the Hall bar in Fig7c. The Hall mobility is extracted as $\mu_H = \sigma/ne$ where $n$ is extracted by measuring the Hall voltage with a magnetic field out of plane B= 0.5 T applied to the sample.

# Crystallinity of the Pt substrate

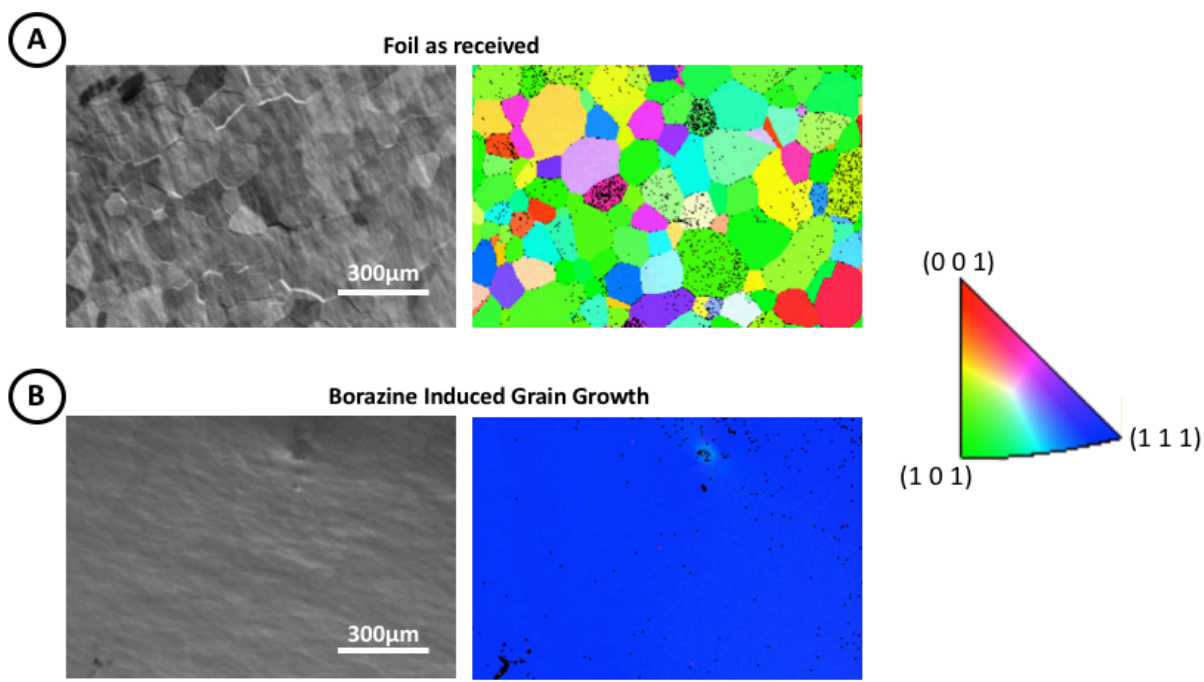

**Figure S 15** EBSD measurement of the catalyst complementary to the XRD results shown in Fig. 3 of the main manuscript. **(A)** SEM image (left) and EBSD (right) image of the same location on a Pt foil as received. The colour indicates the grain orientation. The image clearly shows an plethora of different orientations on the polycrystalline foil. **(B)** SEM image (left) and EBSD (right) image of a foil a Pt foil, which has been subject to borazine annealing ($P_{sc} = 10^{-5}$ mbar, $T_{gr} = 1200$ °C, $t_{an} = 15$ min and. $t_{sd} = 2$ min). The entire imaged area has the Pt (1 1 1) orientation, in agreement with the XRD data given in Fig. 3.

# References

1. Caneva, S. *et al.* Controlling catalyst bulk reservoir effects for monolayer hexagonal boron nitride CVD. *Nano Lett.* **16,** 1250–1261 (2016).

2. Kidambi, P. R. *et al.* In situ observations during chemical vapor deposition of hexagonal boron nitride on polycrystalline copper. *Chem. Mater.* **26,** 6380–6392 (2014).

3. Preobrajenski, A. B., Vinogradov, A. S. & Mårtensson, N. Monolayer of h-BN chemisorbed on Cu(1 1 1) and Ni(1 1 1): The role of the transition metal 3d states. *Surf. Sci.* **582,** 21–30 (2005).

4. Westerström, R., Mikkelsen, A., Lundgren, E. & Preobrajenski, A. A single h-BN layer on Pt (1 1 1). (2008).

5. Caneva, S. *et al.* Nucleation control for large, single crystalline domains of monolayer hexagonal boron nitride via Si-doped Fe catalysts. *Nano Lett.* **15,** 1867–1875 (2015).

6. Trehan, R., Lifshitz, Y. & Rabalais, J. W. Auger and x-ray electron spectroscopy studies of h-BN, c-BN, and $N_2^+$ ion irradiation of boron and boron nitride. *J. Vac. Sci. Technol. A Vacuum, Surfaces, Film.* **8,** 4026–4032 (1990).

7. Ma, T. *et al.* Repeated growth-etching-regrowth for large-Area defect-free single-crystal graphene by chemical vapor deposition. *ACS Nano* **8,** 12806–12813 (2014).

8. Cai, Q. *et al.* Raman signature and phonon dispersion of atomically thin boron nitride. *Nanoscale* **9,** 3059–3067 (2017).